\documentclass{aaprep}

\usepackage{natbib}
\usepackage{graphicx}

\bibpunct{(}{)}{;}{a}{}{,} 

\usepackage{txfonts}

\def\OO#1{{\cal O}(c^{-#1})}
\newcommand{\vecg}[1]{\mbox{\boldmath$#1$}}
\newcommand{\ve}[1]{\vecg{#1}}
\def\muas{\hbox{$\mu$as}}

\def\L{\Lambda}
\def\Lb{\bar{\Lambda}}
\def\g{\gamma}
\def\k{\ve{k}}
\def\Sb{\bar{S}}

\def\ggg{\frac{\gamma^2}{1+\gamma}}
\def\vnun{\hat{\ve{n}}}
\def\vNun{\hat{\ve{N}}}
\def\nun{\hat{n}}

\def\knh{\ve{k}\cdot \hat{\ve{n}}}

\def\sig{\sigma}

\def\a{\alpha}

\def\3{\ss }

\def\vN{{\bf N}}

\def\bdis{\begin{displaymath}}
\def\edis{\end{displaymath}}

\begin{document}

\title{Relativistic effects on the imaging by a rotating optical system}

\author{Guillem Anglada \inst{1} \and Sergei A. Klioner \inst{2}
\and Michael Soffel \inst{2} \and Jordi Torra \inst{1}}

\institute{
Departament d'Astronomia i Meteorologia, Universitat de Barcelona, Av.
Diagonal 647, 08028 Barcelona, Spain
\and
Lohrmann Observatory, Dresden Technical University,
Mommsenstr. 13, 01062 Dresden, Germany}

\offprints{Guillem Anglada, \email{anglada@am.ub.es}}

\date{Received \today  / Accepted \today}

\abstract
{High accuracy astrometric instruments like Gaia aiming at an
accuracy of 1 microarcsecond cannot be considered as point-like observers in the
framework of relativistic modelling of observable quantities.}
{Special-relativistic effects on the imaging by a
non-point-like arbitrarily moving optical instrument are discussed.}
{A special-relativistic reflection law for a mirror of arbitrary shape and
motion is derived in the limit of geometrical optics.
The aberration patterns are computed
with ray tracing using a full special-relativistic model
for two simple rotating optical instrument.}
{It was found that the effect of special-relativistic reflection law on
the photocenters of aberration patterns of an optical system rotating
with a moderate angular velocity of $60 \arcsec/{\rm s}$ may be at the
level of 1 microarcsecond if the system involves mirrors significantly
inclined relative to the optical axis.}
{Special-relativistic optical modelling of the future astrometric instrument
is generally speaking indispensable if a level of a few microarcseconds is
envisaged.}


\titlerunning{Relativistic effects for rotating optical systems}
\authorrunning{G.~Anglada et al.}

\maketitle

\section{Introduction}

The purpose of this paper is to investigate possible relativistic
effects on the imaging by an optical system in arbitrary motion.
Normally, in the framework of relativity one considers point-like
observers. The methods to calculate observed quantities for such
observers are well known. It is tacitly assumed herewith that the
actual instrumentation of the observer is so small that one considers
the positions and velocities of each part of the instrument to be the
same (and that single position and velocity is called the position and
velocity of the observer). In reality even for an Earth-based telescope
it is clear that the velocities of different parts of the primary
mirror in inertial coordinates (not rotating with the Earth) are
slightly different. However, in the past the accuracy of observations
was considered to be  ``too low'' and the size of the mirror ``too
small'' for that differences to be of practical relevance.

Due to recent technical developments especially for astrometric space
missions like Gaia
\citep{GAIA:2000,Perryman:et:al:2001,Bienayme:Turon:2002}, JASMINE
\citep{JASMINE:2002} and SIM \citep{SIM:1998} the situation has
changed. In case of Gaia, we deal with a scanning satellite which
permanently rotates in space with a period of 6 hours. The size of the
primary mirror of Gaia is 1.4 m, that is comparable with the size of
the spacecraft itself. The envisaged best accuracy of Gaia is a few \muas\
(and can be even below that limit in some favorable cases). Therefore,
one cannot neglect a priori the difference of velocities of various
parts of the instruments. It is our purpose to investigate these
effects and estimate their magnitude for Gaia.

The general-relativistic model for Gaia has been formulated in full
detail by \citet{Klioner:2003,Klioner:2004}. The model uses two
principal relativistic reference systems: (1) the Barycentric Celestial
Reference System (BCRS) and (2) the Center of Mass Reference System
(CoMRS) of the satellite. The former is a global reference system with
origin at the barycenter of the solar system. It has been recommended
by the International Astronomical Union for relativistic modelling of
high-accuracy astronomical observations \citep{Soffel:etal:2003}. This
reference system is used to model the dynamics of massive bodies, space
vehicles (e.g., the Gaia satellite) and light rays within the Solar system.
The final Gaia catalogue will contain coordinates of celestial objects
in the BCRS. The CoMRS is the local relativistic reference system of
the satellite. It is explicitly constructed by \citet{Klioner:2004}.
The gravitational influence of massive bodies is effaced in the CoMRS
as much as possible and, according to the equivalence principle,
represented by tidal potentials. The CoMRS has its origin in the center
of mass of the satellite and is kinematically non-rotating with respect
to the BCRS. The CoMRS is physically adequate to model phenomena
occurring in the immediate neighborhood of the satellite: attitude, the
process of observation, etc. According to \citet{Klioner:2004} the
metric tensor of the CoMRS differs from the Minkowski metric by three
kinds of terms (the gravitational field of the satellite is too small
and can be neglected safely): an inertial term due to non-gravitational
accelerations of the satellite (for Gaia these accelerations can be
relatively large during orbital maneuvers and only about
$2\times10^{-13}\,{\rm m}/{\rm s}^2$ in between mainly due to solar
pressure); an inertial term due to the slow rotation of the CoMRS relative
to the co-moving  Fermi-Walker transported locally inertial reference
system (with angular velocity of $\sim 3\times10^{-15}\,{\rm
s}^{-1}=2\,\arcsec\,{\rm per\ century}$); and tidal gravitational
potentials (producing relative accelerations of at most $10^{-12}\,{\rm
m}/{\rm s}^2$ at a distance of 2.5 meters from the satellite's center
of mass). Simple calculations show that all these terms influence the
CoMRS light propagation within a few meters from the satellite's center
of mass at a level much lower than the goal accuracy of 1~\muas.
Therefore, all these terms can be neglected for our purposes and one
can consider the CoMRS for a sufficiently small interval of time
as an inertial reference system of Special
Relativity.

In Section \ref{Section:Idea} we summarize the ideas of how to
calculate the special-relativistic effects in the aberration patterns
due to the rotation of the instrument. Section \ref{sec:imaging} is
devoted to a description of ray tracing calculations of the
relativistic effects in the aberration patterns for two simple optical
systems. The details of the derivation of the special-relativistic
deflection law are given in the Appendix. There we also introduce a
general theoretical scheme which we use to treat arbitrarily-shaped and
arbitrarily moving mirrors in special relativity.

\section{General scheme of computing relativistic effects due to
the rotation of an optical system}
\label{Section:Idea}

Our general goal is to discuss and calculate the influence of
relativistic effects on the imaging by an optical instrument in some
non-inertial motion. In this paper we simplify our general goal in
several directions: (1) we consider here the case of optical
instruments consisting of mirrors only (no lenses are considered), (2) we
do not consider the effects of wave optics and work in the
approximation of geometric optics (see, however, a note at the end of
Section \ref{Section:conclusion}).

For an optical system consisting solely of a number of arbitrarily
moving mirrors, the most important relativistic effect is the
special-relativistic modification of the reflection law. That modified
special-relativistic reflection law will produce a change in aberration
patterns as compared to the patterns calculated by using the normal Newtonian
reflection law. These perturbed aberration patterns could potentially
affect astrometric measurements based on an interpretation of the
images obtained in the instrument's focal plane.

\subsection{Reflection law}

First, we have to formulate the general principles allowing one to
calculate the aberration patterns within the framework of Special
Relativity. Let us consider the following problem within the framework
of Special Relativity. Given a mirror of arbitrary shape in arbitrary
motion
(see Section \ref{sec:mirror} for a formal mathematical description of
such an arbitrary mirror and Section \ref{Section:physical_mirrors}
for a discussion of such mirrors from the physical point of view)
and a light ray hitting the surface of the mirror at a given
point and moment of time we would like to calculate the parameters of
the outgoing (reflected) light ray. The simplified problem of a plain mirror moving
with a constant velocity perpendicular to its surface has been
considered by \citet{Einstein:1905} in the very first paper on Special
Relativity Theory. In the Appendix the most general case of this
problem within Special Relativity is considered in great detail.
Slightly modifying the arguments of \citet{Einstein:1905} we first use
Lorentz transformations to transform from a laboratory inertial
reference system $(t,x^i)$ to an inertial reference system $(T,X^a)$
instantaneously co-moving with the element of the mirror where the
reflection of a particular light ray occurs, then apply the known
reflection law in that reference system and then transform the
reflected light ray back into the laboratory reference system. The
relation of that scheme to direct calculations involving Maxwell's equations
is also discussed in the Appendix. In our calculations we recover a number of
known results for various particular cases. An overview of these known
results and the corresponding comparison are also given. The main
formula used in all the ray tracing calculations of Section~\ref{sec:imaging}
is the relativistic reflection law given by Eq. (\ref{sigma-prime}).

\subsection{Arbitrarily shaped and moving mirrors}
\label{Section:physical_mirrors}

A very important point of the whole scheme is that the shapes of the
mirrors in laboratory coordinates $(t,x^i)$ and, possibly, the
time-dependence of these shapes are assumed to be {\it given}. We describe
the shape of each mirror by a two-parameter family of worldlines of
each individual particle of the mirror denoted as $x_m^i(t;\xi,\eta)$.
Here $\xi$ and $\eta$ are two continuous parameters ``numbering'' the
particles which constitute the surface of the mirror. Clearly, for
fixed values of $\xi$ and $\eta$ function $x^i_m(t;\xi,\eta)$
represents the $(t,x^i)$-parametrization of the world line of the
corresponding particle. For fixed $t$ the same function
$x^i_m(t;\xi,\eta)$ represents the instantaneous position and shape of the
mirror in the $t={\rm const}$ hyperplane of the coordinates $(t,x^i)$.
In this case ($t={\rm const}$) the parameters $\xi$ and $\eta$ give a
kind of non-degenerated two-dimensional coordinate chart on the surface
of the mirror. We consider $x^i_m(t;\xi,\eta)$ to be differentiable
with respect to $\xi$ and $\eta$. This means that the coordinate
representation of the surface is a smooth two-dimensional surface for
each moment of coordinate time $t$.

Let us note that in general there is no inertial reference system where
the whole system or any of its mirrors is at rest. In the special
cases when such an inertial rest-frame of a mirror does exist one
should certainly consider the shape of the mirror in that rest-frame.
In the practical cases considered below such rest-frames do not exist.
Moreover, the size of the mirrors is so large that we cannot assume
that the velocities of all points of the mirror are approximately
constant in any inertial reference system.

We do not consider the question of deformations of the mirrors due to
their non-inertial (for example, rotational) motion (i.e., the relation
between the intended shapes of the mirrors during their manufacturing
and their shapes, e.g., in a rotating satellite, in coordinates
$(t,x^i)$). The behaviour of a mirror as a physical body is a separate
question, a rigorous relativistic treatment of which would require at
least a special-relativistic theory of elasticity. As long as the
angular velocity is constant the deformations and special-relativistic
effects on the shape (e.g. Lorentz contraction) are also constant. In
this case a rigidly rotating mirror can be considered to be Born-rigid
\citep[Section 45]{pauli:1958}. We can also argue that the constant
deformations are assumed to be properly taken into account during
manufacturing so that the rotating mirrors have the assumed forms. One
can even argue that the mirrors can be made active to retain the
prescribed form (which is certainly the case for many larger
Earth-bound instruments, but may appear to be a rather bizarre argument
in some other cases).

\subsection{``Observable'' aberration patterns}

The last issue is the definition of the observing (imaging) device. In
analogy to our representation of the mirrors we first define a
coordinate ``plane'' $x_f^i(t;\zeta,\chi)$ in laboratory coordinates
$(t,x^i)$ which coincides with the focal ``plane'' of the instrument in
the Newtonian case. In many cases (e.g. for the case considered in
Section \ref{sec:imaging} below) $x_f^i(t;\zeta,\chi)$ can be taken to
be a plane in the considered coordinates (that is, for any moment of
time there exist $n^i(t)$ independent of $\zeta$ and $\chi$ such that
$\ve{x}_f\cdot\ve{n}=0$). The aberration patterns we calculate below
are defined as the set of points at which the light rays from a source
hit that coordinate focal plane at some moment $t=t_{\rm obs}={\rm
const}$. Generally speaking the aberration patterns cannot be
considered as ``infinitely small''. This means that there is no
inertial coordinate system in which the part of the detector (that is,
of the focal ``plane'') registering an aberration pattern can be
considered at rest.

If the patterns are ``small enough'' (which is typical case for
reasonable high-quality optical instruments) one could introduce an
inertial reference system $(\tau,\rho^i)$ instantaneously co-moving
some central point of the aberration pattern and define the
``observable'' pattern as a set of points at which the light rays from
a source hit that coordinate focal plane at some moment $\tau=\tau_{\rm
obs}={\rm const}$ (here one should also take into account the
relativistic effects in spatial coordinates and correspondingly treat
Lorentz contraction etc.). First, although this approach seems to be
more adequate for a non-inertial motion it still gives a
coordinate-dependent picture because of finite extension of the
patterns. Second, we have explicitly checked that this additional
Lorentz boost does not influence any of the pictures and numerical
results given below.

Note that we are interesting in prediction of the changes in the
aberration patterns compared to the prediction made for the ``same''
optical device without rotation and using Newtonian geometric optics
(this latter prediction is typically available from the
manufacturers of the instrumentation). From this point of view, our
definition of ``observed'' aberration pattern is adequate. In more
realistic case one has to model the process of observation in much more
detail (e.g., CCD orientation and position within the instrument, CCD
clocking, averaging, TDI mode etc.). Such a detailed modelling is however
unnecessary for the purposes of this paper.

\bigskip

Summarizing, our aberration pattern modelling consists of (1) fixing
the models of the mirrors $x^i_m(t;\xi,\eta)$ and the focal plane
$x_f^i(t;\zeta,\chi)$, and (2) tracing a grid of incoming light rays,
which interact with the optical system only at the moments of
reflection according to (\ref{sigma-prime}), until the point of
intersection with the focal plane $x_f^i(t;\zeta,\chi)$, and (3)
forming the aberration pattern itself and/or calculating its
photocenter.

\section{Relativistic astrometric effects due to rotational motion of the
satellite}
\label{sec:imaging}

In order to evaluate the relativistic effects in the aberration
patterns of planned scanning astrometric instruments, let us consider
an extended optical system rotating rigidly with a constant angular
velocity relative to the inertial reference system $(t,x^i)$. For a
scanning astrometric satellite the real angular velocity is not
constant (e.g., because of the required scanning law), but its changes
are small and slow, and will be neglected here. Rigid rotation of the
optical instrument means that the whole instrument is at rest in a
reference system $(t,y^i)$ related to the inertial laboratory reference
system $(t,x^i)$ as $y^i=R^i_{\ j}\,x^j$, $R^i_{\ j}$ being an
orthogonal (rotational) matrix.

To calculate the aberration patterns of several optical systems
discussed below we have developed a numerical ray tracing code in Java
allowing us to calculate aberration patterns for an arbitrary optical
system rigidly rotating in our laboratory coordinates. Each mirror in
the system can be individually shaped and oriented in those
coordinates. The code allows us to control all intermediate
calculations as well as the overall numerical accuracy.

Parameters of the optical systems (size of the mirrors, focal distance,
distance of the primary mirror from the rotational axis and angular
velocity) considered in Sections \ref{sec:single} and
\ref{sec:realdevice} below are chosen to represent qualitatively some
principal features of planned astrometric missions like Gaia
\citep{Perryman:et:al:2001} or JASMINE \citep{JASMINE:2002}, where a
scanning satellite comprising two astrometric telescopes continuously rotates with
an angular velocity of order $\Omega \sim 60\ \arcsec/{\rm s}$.

\begin{figure}[htb]
\centering
\includegraphics[width=8.5cm]{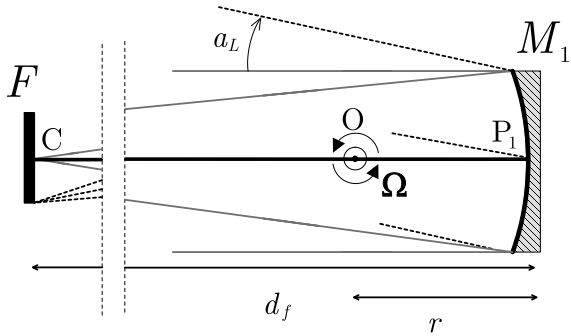}
\caption{A rotating optical system with one mirror. The primary
mirror $M_1$ is parabolic. The point $\mathrm{P}_1$ lies on the
vertex of the parabola. The distance from the origin $\mathrm{O}$ to
$\mathrm{P}_1$ is $r$. The distance from $\mathrm{P}_1$ to the focal
plane center $\mathrm{C}$ is the focal distance of the parabola
$d_f$. The optical system rotates rigidly around the origin
$\mathrm{O}$ with an angular velocity $\Omega$ in the sense shown on
the scheme. The direction of the incoming light ray is parameterized
with two angles: the \textit{along scan} angle $a_L$ in the plane of
the depicted scheme (this angle is changing continuously for a given
source because of the rotation), and the across scan angle $a_C$
(not shown in the picture) in perpendicular direction. The
instantaneous optical axis is represented by the bold horizontal
line going from $\mathrm{P}_1$ to $C$. Without rotation light rays
parallel to the optical axis converge to the single point $C$ in the
focal plane. \label{fig:singlemirror} }
\end{figure}

\subsection{A One-mirror optical system}
\label{sec:single}
\label{sec:one-mirror}

The first optical system that we will study consists of one rotating
parabolic mirror. A scheme of this optical system is given on
Fig.~\ref{fig:singlemirror}. The parabolic mirror $M_1$ is a square
mirror of size 1.5~m $\times$ 1.5~m and focal length $d_f =
46.67$~m. This roughly corresponds to the astrometric instruments of
Gaia. The receiver at the focal plane is considered to be 0.814~m
$\times$ 0.814~m in size providing a field of view of $\sim 1\degr\times
1\degr$. The rotational axis goes through the origin $\mathrm{O}$ of
our coordinates perpendicular to the plane of
Fig.~\ref{fig:singlemirror}. The distance from $\mathrm{O}$ to the
center of the primary mirror $\mathrm{P}_1$ is $r = 1.5$~m. The
distance from $\mathrm{P}_1$ to the center of the focal plane
$\mathrm{C}$ is obviously the focal distance $46.67$~m. The whole
optical system is rotating with respect to $\mathrm{O}$ with an
angular velocity $\Omega = 60\ \arcsec/{\rm s}$. The \emph{optical
axis} of the system is defined as the path of the light ray which
goes perpendicular to the surface of the primary mirror through its
center provided that the system does not rotate. The direction of an
incoming light ray is parameterized with two angles: the
\textit{along scan} angle $a_L$ and the \textit{across scan} angle
$a_C$ (see Fig.~\ref{fig:singlemirror}). The along scan angle is the
angle between the instantaneous directions of the optical axis and
the incoming light ray projected into the plane containing the
optical axis and perpendicular to the vector of angular velocity of the
system (i.e., the plane of Fig.~\ref{fig:singlemirror}).
The across scan angle is the angle between the instantaneous
directions of the optical axis and the incoming light ray projected
into the plane containing both the optical axis and the vector of
angular velocity. The along scan and across scan angles are widely
used in the context of scanning astrometric missions like HIPPARCOS
\citep{HIPPARCOS:1997} and Gaia \citep{Perryman:et:al:2001}.

\begin{figure}
\centering

\includegraphics[width=200pt,clip=true]{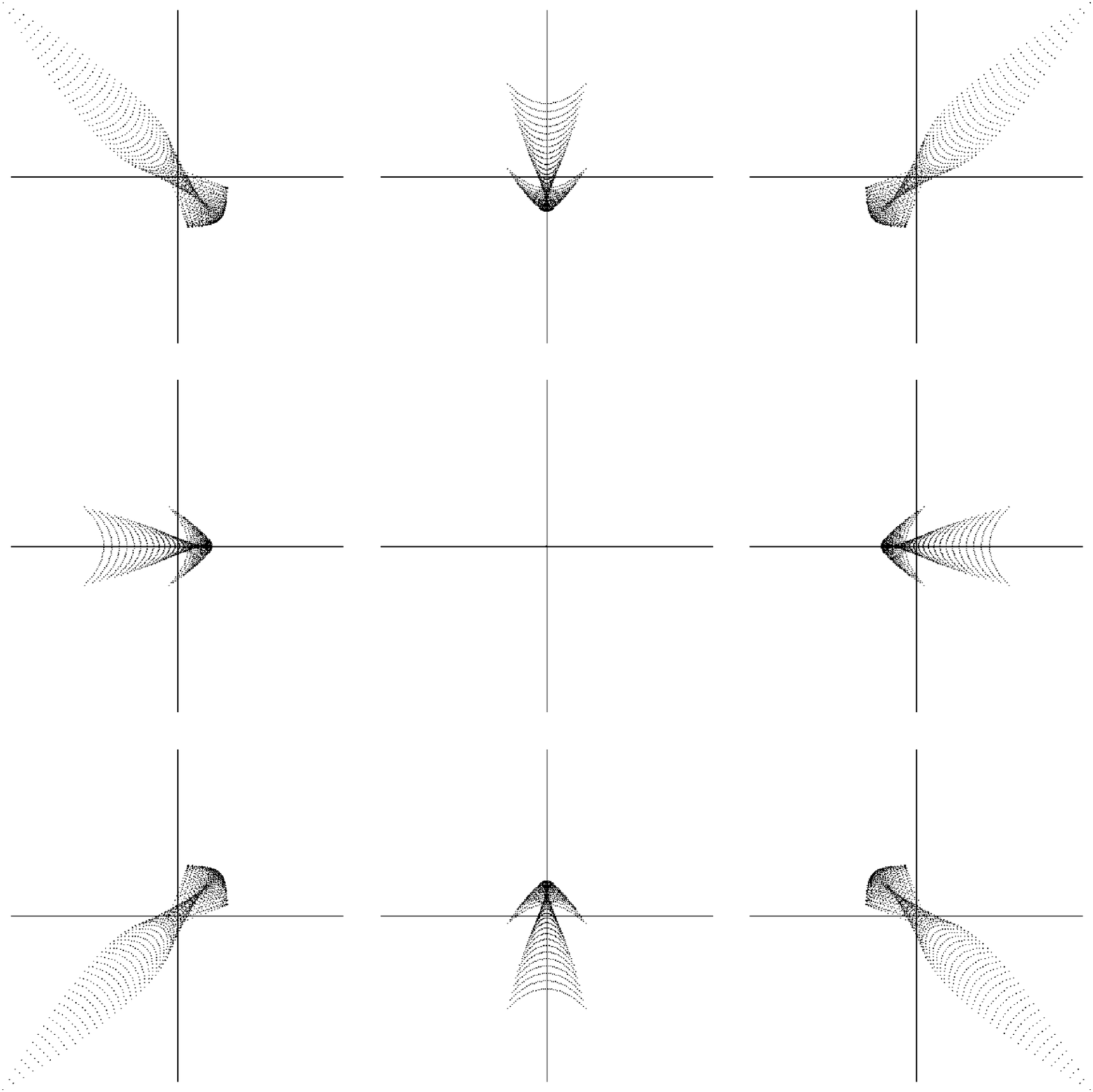}

\vskip -12pt

\underline{\hbox to 8cm{}}

\vskip -2pt

\includegraphics[width=200pt,clip=true]{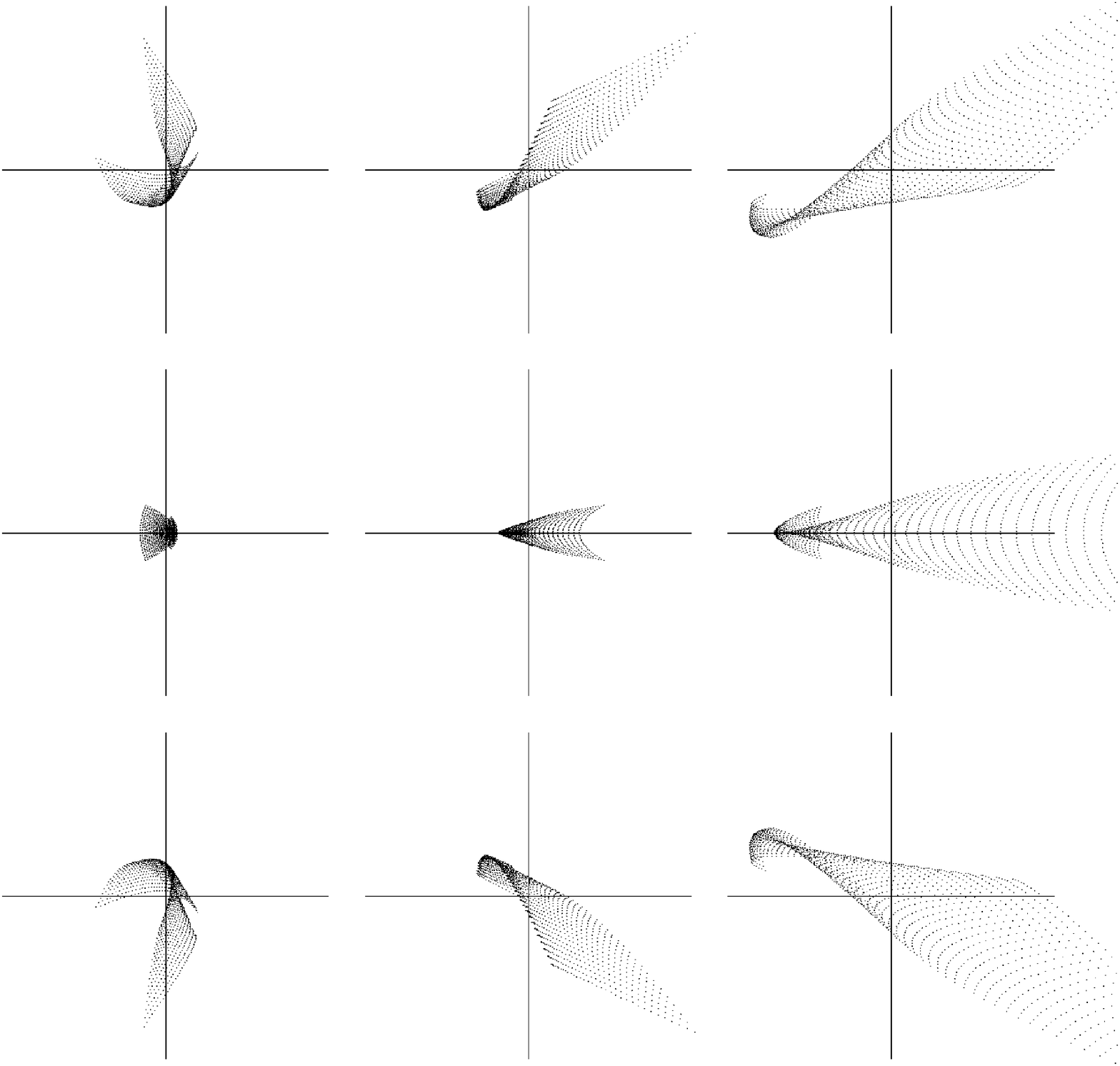}

\vskip -12pt

\underline{\hbox to 8cm{}}

\vskip -2pt

\includegraphics[width=200pt,clip=true]{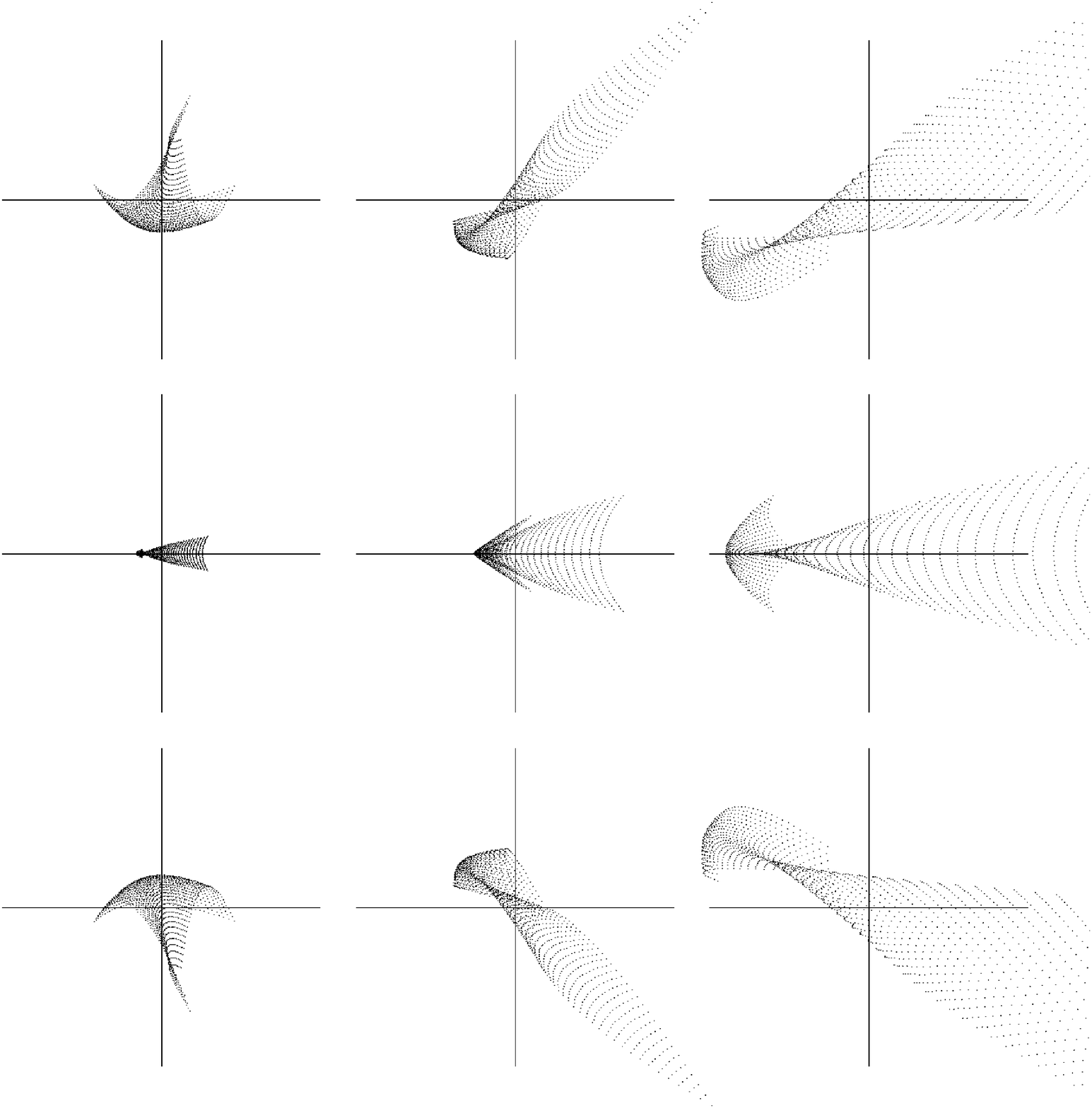}

\vskip -2pt

\caption{Aberration patterns for the one-mirror system: non-rotating
instrument (upper pane), rotating instrument considering the
light propagation delays and using the Newtonian reflection law (middle
pane), and rotating instrument considering both the light propagation
delays and the relativistic reflection law (lower pane). An extremely high
angular velocity $\Omega = 5\times 10^9 \arcsec/{\rm s}$ is used in
order to and make the distortion clearly visible. See text for further
explanations.
\label{fig:single_patterns} }
\end{figure}

In order to evaluate the effects due to the rotation of the instrument
we calculate aberration patterns for different values of the field
angles $a_L$ and $a_C$ as well as the differences of the photocenters
for each considered case. To compute aberration patterns a rectangular
grid of parallel incoming light rays with direction characterized by
some given $a_L$ and $a_C$ is generated. These light rays are then
traced through the optical system until they intersect the focal plane.
The coordinates of the intersection points produce the corresponding
aberration pattern in the focal plane (see, e.g., Figs.
\ref{fig:single_patterns} and Fig.\ref{fig:patterns}). The photocenter
of a pattern is defined as the mean position of all points of that
pattern.

We distinguish between two different effects changing the aberration
patterns (and their photocenters) of a rotating instrument compared to
those of some identical non-rotating instrument. The first effect is the change of
orientation of various reflecting surfaces during the time delays
needed for a light ray to propagate from the primary mirror to the focal plane.
The second effect is the difference between
the Newtonian and relativistic reflection laws.

Clearly, the propagation delays are related just to the finiteness of
the light velocity. The delays appear also in the non-rotating case,
but can be completely ignored since the orientation of all reflecting
surfaces is constant. For a rotating instrument the propagation delays
mean in particular that the light rays producing an aberration pattern
(that is, the light rays intersecting the focal plane at the same
moment of time) hit the primary mirror (and, generally speaking, all
other mirrors) at different moments of time. The effect of propagation
delays can be directly calculated in our ray tracing software by using
a specially designed iterative scheme.

There are several effects related to the propagation delays, some of
which are quite easy to understand. The first one is just the constant
shift of the aberration patterns due to the change of the orientation
of the instrument during the propagation time: an image of a star
observed at time $t_{\rm obs}$ is produced by the light rays from the
star which hit the primary mirror at time $\sim\,t_{\rm obs}-d_f/c$
when the orientation of the latter differed by $\sim\,\Omega\,d_f/c$
from the orientation at $t_{\rm obs}$. Similar constant shifts will be
caused by intermediate mirrors and by the motion of the focal plane
during the propagation delay: during the light propagation the focal
plane is moving and the photon hits the focal plane at different
position which corresponds to a different position on the sky. It is
clear that the latter effect can be computed as
$\sim\,\Omega\,(d_f-r)/c$ for the one-mirror system depicted in
Fig.~\ref{fig:singlemirror}. Note that in the limit when the center of
rotation is infinitely far from the instrument (that is, when all parts
of the instrument effectively have the same velocity), these constant
shifts are fully equivalent to normal aberration of light. The constant
shifts of the aberration patterns, which could be relatively large,
effectively lead only to a constant time shift in the orientation
parameters of the satellite derived from astrometric observations: the
orientation obtained  from observations at $t_{\rm obs}$ is actually
the orientation the satellite had some small time interval earlier.
This hardly has consequences on the measurements in any existing or
planned astrometric projects. However, the propagation delays lead also
to a deformation of aberration patterns which depends on the field
angles. Those aberration pattern deformations together with the
deformations due to the relativistic reflection law can be important as
illustrated below. The distortions of the shape of the patterns are
caused by different velocities of different parts of both mirrors and
slightly different incident angles for each mirror.

For the one-mirror case these effects are illustrated in
Fig.~\ref{fig:single_patterns}. The nine patterns in each of the three
panes correspond to nine combinations of the field angles with
$a_L=-30\arcmin, 0\arcmin, +30\arcmin$ (horizontal direction) and
$a_C=-30\arcmin, 0\arcmin, +30\arcmin$ (vertical direction). For the
focal length $d_f=46.67$ m, $30 \arcmin$ corresponds to about 407 mm in
the focal plane coordinates. The size of the axes in focal plane
coordinates is 0.5~mm $\times$ 0.5~mm for all patterns. The aberration
patterns on the upper pane are calculated for a non-rotating
instrument. On the middle pane the aberration patterns are obtained
using the Newtonian reflection law, but the effects of the light
propagation delays are taken into account. On the lower pane both the
light propagation delays and the relativistic reflection law are used.
An extremely high angular velocity $\Omega = 5\times 10^9 \arcsec/{\rm
s}$ is used in order to exaggerate the distortion and make it clearly
visible. The three rightmost patterns on
both the middle and the lower panes are much larger than all other
patterns. These six patterns extend to the left from the edge of Figure
by about 3 times the size of the horizontal axis in each pattern. These
parts of the patterns are not shown in Fig.~\ref{fig:single_patterns}.
The axes for each pattern are centered at the corresponding
photocenter. Note that those photocenters are significantly shifted
between the three panes due to the constant propagation time effects
discussed above.

Since for the one-mirror instrument
the angle of each light ray with respect to the normal to the mirror at
each point of the surface is not greater than $30\arcmin$, the effect
of the relativistic reflection law on aberration patterns for the
one-mirror system is very small. At point $\mathrm{P}_1$ the velocity
vector is perpendicular to the normal to the mirror. Therefore, at this
point for any $a_L$ and $a_C$ the relativistic reflection law coincides
with the Newtonian one (see Eq. (\ref{sigma-prime})). A light ray going
through that point will intersect the focal plane in the same point for
both Newtonian and relativistic reflection laws. The light rays of the
same grid not going through $\mathrm{P}_1$ have different images in the
Newtonian and relativistic cases.

For realistic $\Omega=60\ \arcsec/{\rm s}$ the mean shift of the
photocenters due to the propagation delays amount to $\delta
\overline{a}^{\,d}_L = 18.3842$~\muas. Note that this number can be
reproduced with a good accuracy by $\Omega\,(2d_f-r)/c=18.3807$~\muas\
as discussed above. The field-angle dependent change of the
photocenters is at the level of 0.001~\muas\ and is shown in Table
\ref{Table-one-mirror}. The change of the photocenters due to
relativistic reflection law turns out to be a shift in the along-scan
direction $\delta a_L\approx\delta \overline{a}^{\,r}_L =
-0.0008$~\muas\ and is independent of $a_L$ and $a_C$ at the level of
0.0001~\muas.

\begin {table}
\center
\begin{tabular}{r|rrr|rrr}
                   & \multicolumn{3}{c|}{$\delta a_L\ \times 10^{-3} \muas$}                         & \multicolumn{3}{c}{$\delta a_C\ \times 10^{-3} \muas$} \\[1mm]
 $_{\displaystyle{a_C}}$ $\backslash$ $^{\displaystyle{a_L}}$      & $-30\arcmin$  &  $0\arcmin$   & $+30\arcmin$   &  $-30\arcmin$ & $0\arcmin$ & $+30\arcmin$ \\[1mm]
\hline
&&&&&&\\[-2mm]
$-30\arcmin$\phantom{$\backslash$ $^{\displaystyle{a_C}}$}  &  0.9 & -1.2 &  0.9 &  1.4 & 0.0 & -1.4 \\
  $0\arcmin$\phantom{$\backslash$ $^{\displaystyle{a_C}}$}  &  0.2 & -1.9 &  0.2 &  0.0 & 0.0 &  0.0 \\
 $30\arcmin$\phantom{$\backslash$ $^{\displaystyle{a_C}}$}  &  0.9 & -1.2 &  0.9 & -1.4 & 0.0 &  1.4 \\
\end{tabular}
\caption{The one-mirror optical system rotating at $\Omega=60\ \arcsec/{\rm s}$:
the part of the shifts of the photocenters which depends on
the field angles (that is, the constant shift of
$\delta \overline{a}^{\,d}_L+\delta \overline{a}^{\,r}_L=18.3834$ \muas\
is removed; see text for further explanations).
\label{Table-one-mirror}
}
\end{table}

\subsection{A two-mirror optical system}
\label{sec:realdevice}

\begin{figure}
\centering
\includegraphics[width=8.0cm]{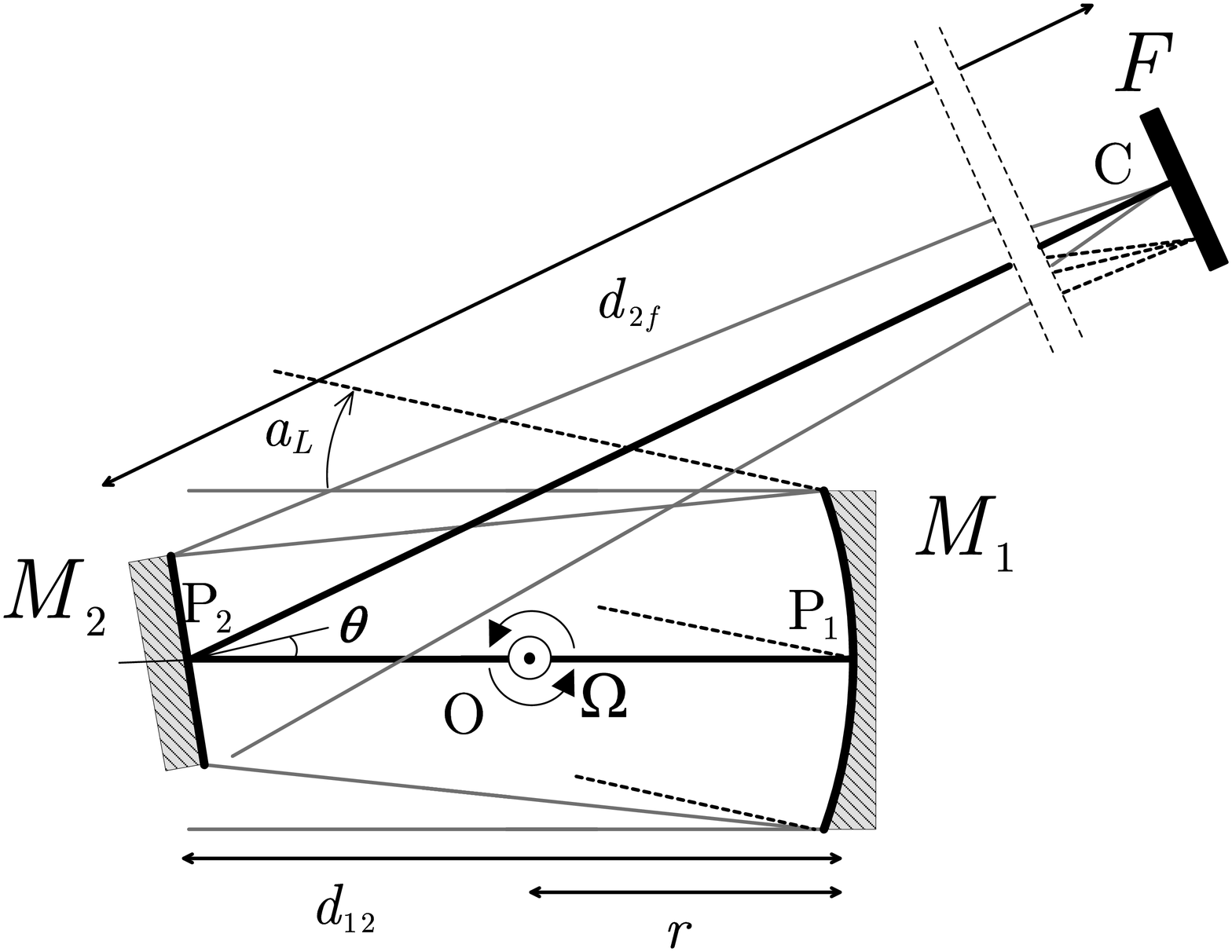}
\caption{A flat secondary mirror $M_2$ has
been added to the optical system depicted on Fig.~\ref{fig:singlemirror}. The
distance from $\mathrm{P}_1$ to center of the flat mirror
$\mathrm{P}_2$ is $d_{12}$. The focal plane position depends on the
angle $\theta$. Now the bold line representing the optical axis
goes from $\mathrm{P}_1$ to $\mathrm{P}_2$ and then to
the focal plane center $C$.
}
\label{fig:device}
\end{figure}

Real optical systems normally have more than one mirror. Often the
instruments involve mirrors inclined by about $45\degr$ to the optical
axis (i.e., Nasmith focus, beam combiners, beam
splitters, etc.). In this case the effects of the relativistic reflection law on the
aberration pattern are significantly larger than in the case discussed
above. Here we consider an optical system consisting of one parabolic
primary mirror and one flat secondary mirror as depicted on
Fig.~\ref{fig:device}. The whole system is again rigidly rotating with
a constant angular velocity $\Omega$ in laboratory coordinates.
The flat mirror is inclined at an angle $\theta$ with respect to the
optical axis of the primary mirror. The distance from $\mathrm{P}_1$ to
$\mathrm{P}_2$ is $d_{12}=3$~m, and the distance from $\mathrm{P}_1$ to
the rotational axis $\mathrm{O}$ is $r=1.5$~m. The distance from
$\mathrm{P}_2$ to the center $C$ of the focal plane is $d_f - d_{12} =
d_{2f}=43.67$~m.

\begin{figure}
\centering

\includegraphics[width=200pt,clip=true]{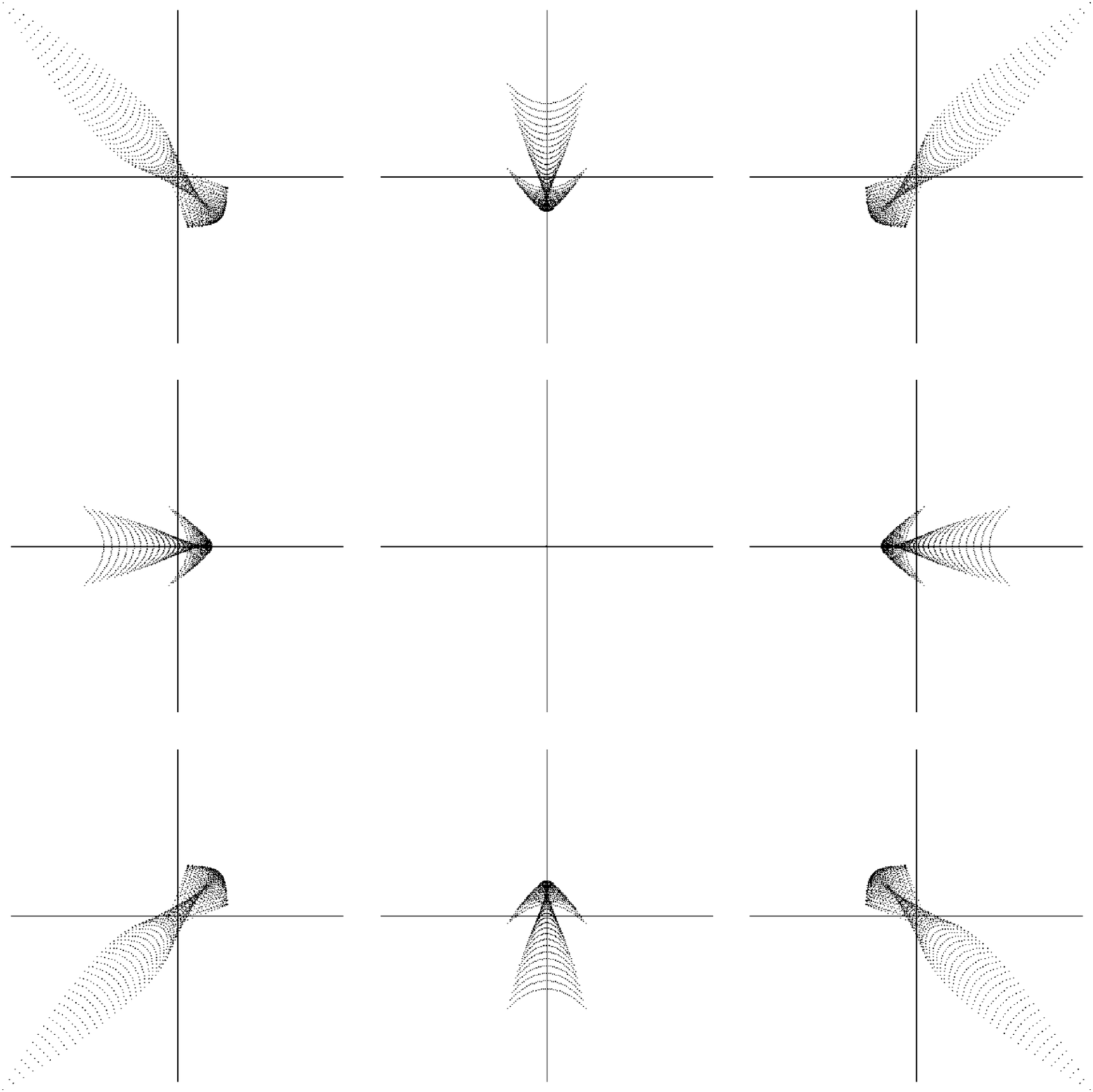}

\vskip -12pt

\underline{\hbox to 8cm{}}

\vskip -2pt

\includegraphics[width=200pt,clip=true]{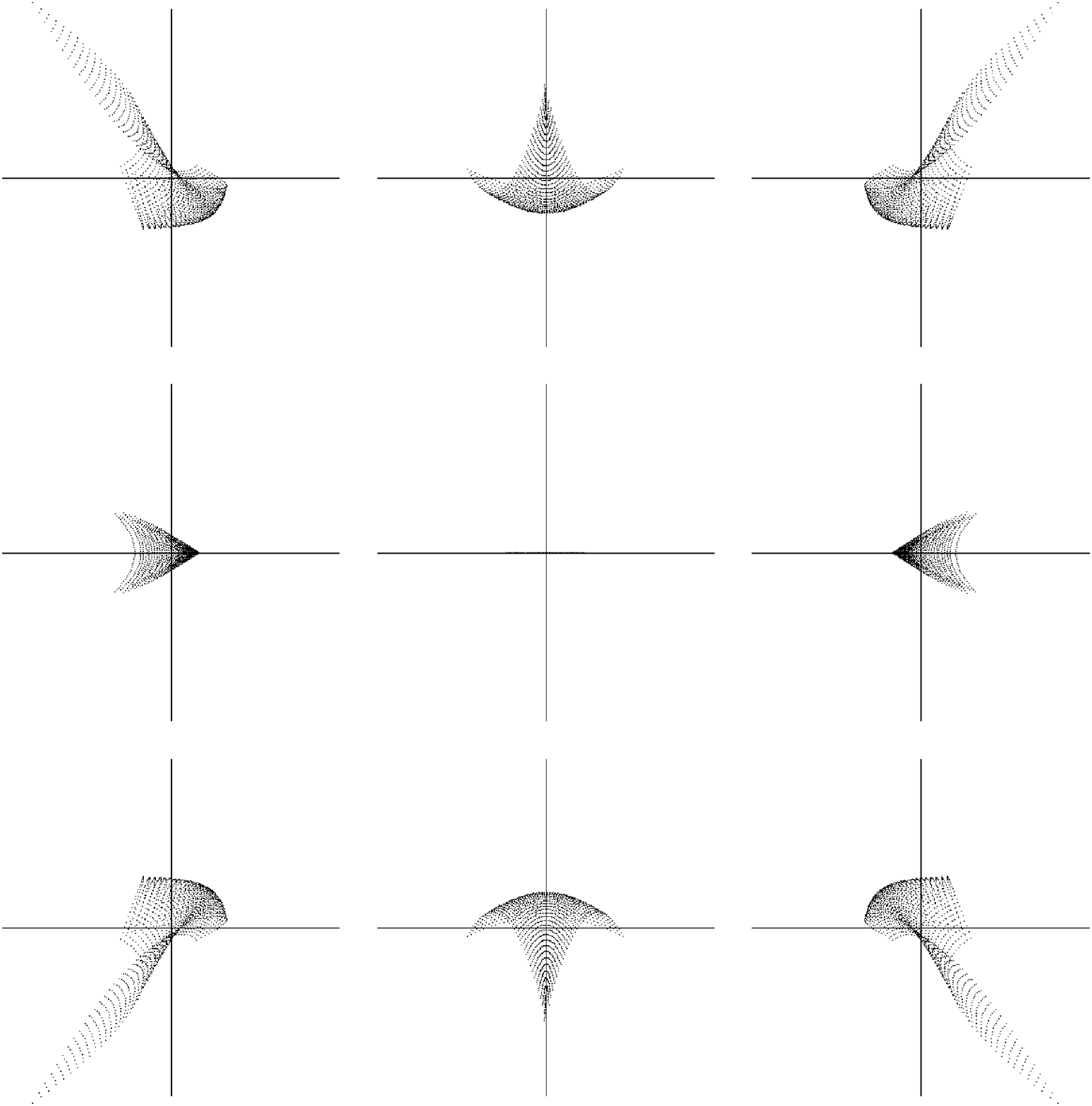}

\vskip -12pt

\underline{\hbox to 8cm{}}

\vskip -2pt

\includegraphics[width=200pt,clip=true]{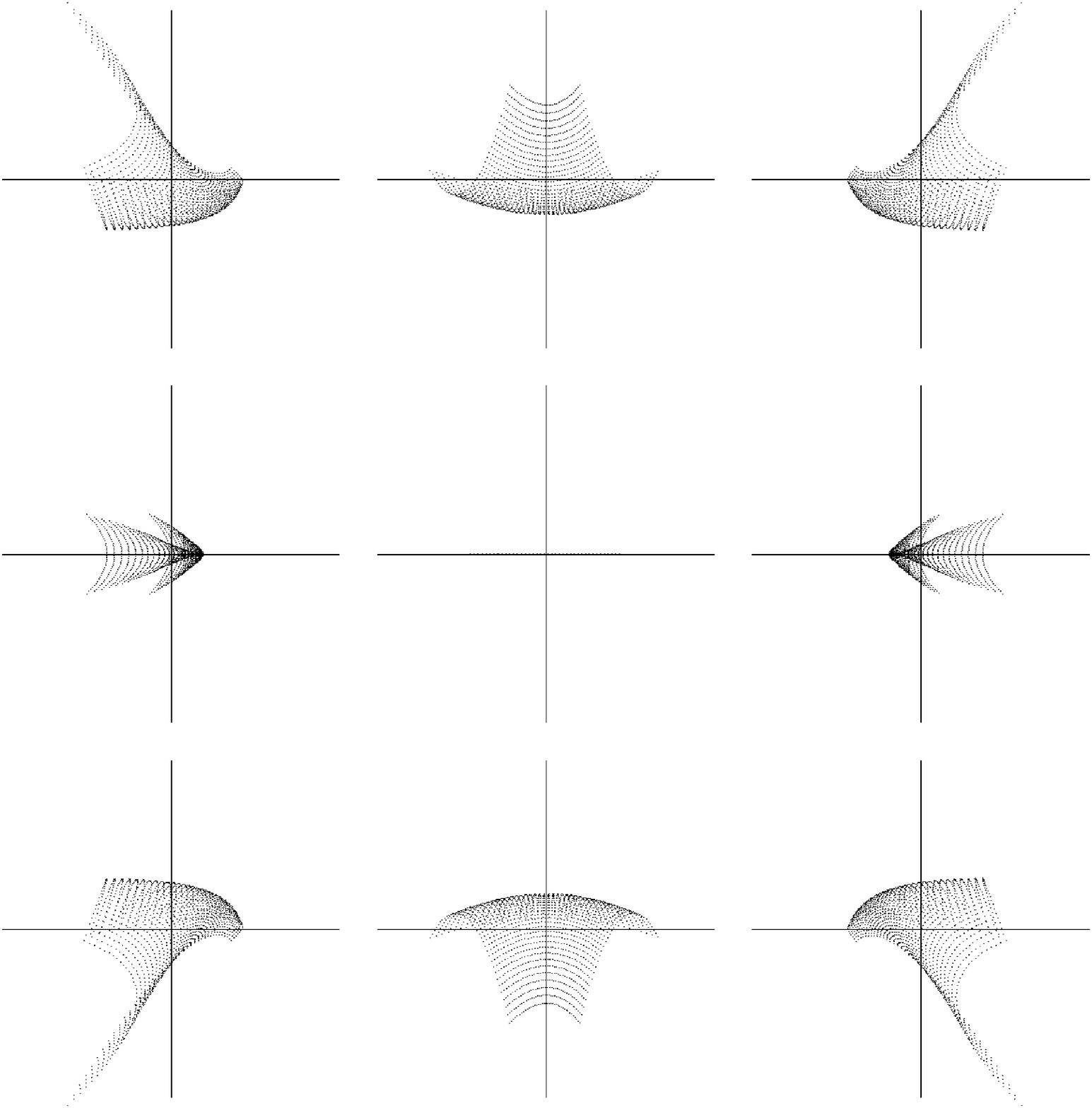}

\vskip -2pt

\caption{
Aberration patterns for the two-mirror system with $\theta = 45\degr$: non-rotating
instrument (upper pane), rotating instrument considering the
light propagation delays and using the Newtonian reflection law (middle
pane), and rotating instrument considering both the light propagation
delays and the relativistic reflection law (lower pane). An
angular velocity $\Omega = 5\times 10^7 \arcsec/{\rm s}$ is used
to make the distortion clearly visible. See text for further
explanations.
\label{fig:patterns}
}
\end{figure}

We repeat the ray tracing calculations as described in
Section~\ref{sec:single} above with this additional flat mirror. We use
three different configurations of the flat mirror with inclination
angles $\theta=+45 \degr$, $\theta=0$, and $\theta=-45 \degr$.
Figure~\ref{fig:patterns} shows the aberration patterns obtained with
$\theta = 45 \degr$ (again for a large angular velocity of $\Omega =
5\times 10^7 \arcsec/\rm{s}$ (100 times lower than for
Fig.~\ref{fig:single_patterns}) was used in order to make the effects
visible). The same 9 combinations of $a_L$ and $a_C$, and the same size
and centering of the axes are used for each pane as described above for
Fig.~\ref{fig:single_patterns}. The upper pane is again for the
aberration patterns for a non-rotating instrument ($\Omega=0$). These
patterns are identical to those on the left pane of
Fig.~\ref{fig:singlemirror}. Clearly, the aberration patterns for the
rotating instrument (the middle and the lower pane) look differently
compared to Fig.~\ref{fig:single_patterns}. Numerical values of the
shifts of the photocenters $\delta a_L$ and $\delta a_C$ for
$\Omega=60\ \arcsec/{\rm s}$ are presented in Table~\ref{tab:numbers}.

\begin {table}[htb]
\center

\begin{tabular}{r|rrr|rrr}
\multicolumn{7}{l}{$\theta=-45\degr:\quad \delta\overline{a}_L^{\,d}=1.7422\,\muas,\quad\delta\overline{a}_L^{\,r}=-0.2776\,\muas\quad$}\\[2mm]
& \multicolumn{3}{|c}{$\delta a_L\ \times 10^{-3} \muas$}             & \multicolumn{3}{|c}{$\delta a_C\ \times 10^{-3} \muas$} \\[1mm]
$_{\displaystyle{a_C}}$ $\backslash$ $^{\displaystyle{a_L}}$ & $-30\arcmin$  &  $0\arcmin$   & $+30\arcmin$   &  $-30\arcmin$ & $0\arcmin$ & $+30\arcmin$ \\[1mm]
\hline
&&&&&&\\[-2mm]
$-30\arcmin $\phantom{$\backslash$ $^{\displaystyle{a_C}}$}  & $ 4.6 $ & $  0.0 $ & $ -4.5 $ & $ 4.8 $ & $ 4.9 $ & $ 4.9 $ \\
$0\arcmin   $\phantom{$\backslash$ $^{\displaystyle{a_C}}$}  & $ 4.5 $ & $ -0.1 $ & $ -4.6 $ & $ 0.0 $ & $ 0.0 $ & $ 0.0 $ \\
$30\arcmin  $\phantom{$\backslash$ $^{\displaystyle{a_C}}$}  & $ 4.6 $ & $  0.0 $ & $ -4.5 $ & $-4.8 $ & $-4.9 $ & $-4.9 $ \\
\multicolumn{7}{c}{}\\[2mm]
\multicolumn{7}{l}{$\theta=0\degr:\quad \delta\overline{a}_L^{\,d}=2.0246\,\muas,\quad\delta\overline{a}_L^{\,r}=0.0006\,\muas\quad$}\\[2mm]
& \multicolumn{3}{|c}{$\delta a_L\ \times 10^{-3} \muas$}                         & \multicolumn{3}{|c}{$\delta a_C\ \times 10^{-3} \muas$} \\[1mm]
$_{\displaystyle{a_C}}$ $\backslash$ $^{\displaystyle{a_L}}$       & $-30\arcmin$  &  $0\arcmin$   & $+30\arcmin$   &  $-30\arcmin$ & $0\arcmin$ & $+30\arcmin$ \\[1mm]
\hline
&&&&&&\\[-2mm]
$-30\arcmin $\phantom{$\backslash$ $^{\displaystyle{a_C}}$}  & $ 0.0 $ & $ 0.0 $ & $ 0.0 $ & $ 0.0 $ & $0.0 $ & $ 0.0 $ \\
$0\arcmin   $\phantom{$\backslash$ $^{\displaystyle{a_C}}$}  & $ 0.0 $ & $ 0.0 $ & $ 0.0 $ & $ 0.0 $ & $0.0 $ & $ 0.0 $ \\
$30\arcmin  $\phantom{$\backslash$ $^{\displaystyle{a_C}}$}  & $ 0.0 $ & $ 0.0 $ & $ 0.0 $ & $ 0.0 $ & $0.0 $ & $ 0.0 $ \\
\multicolumn{7}{c}{}\\[2mm]
\multicolumn{7}{l}{$\theta=+45\degr:\quad \delta\overline{a}_L^{\,d}=1.7422\,\muas,\quad\delta\overline{a}_L^{\,r}=-0.2776\,\muas\quad$}\\[2mm]
& \multicolumn{3}{|c}{$\delta a_L\ \times 10^{-3} \muas$}                         & \multicolumn{3}{|c}{$\delta a_C\ \times 10^{-3} \muas$} \\[1mm]
$_{\displaystyle{a_C}}$ $\backslash$ $^{\displaystyle{a_L}}$       & $-30\arcmin$  &  $0\arcmin$   & $+30\arcmin$   &  $-30\arcmin$ & $0\arcmin$ & $+30\arcmin$ \\[1mm]
\hline
&&&&&&\\[-2mm]
$-30\arcmin $\phantom{$\backslash$ $^{\displaystyle{a_C}}$}  & $-4.5 $ & $ 0.0 $ & $ 4.6 $ & $-4.9 $ & $-4.9 $ & $-4.8 $ \\
$0\arcmin   $\phantom{$\backslash$ $^{\displaystyle{a_C}}$}  & $-4.6 $ & $-0.1 $ & $ 4.5 $ & $ 0.0 $ & $ 0.0 $ & $ 0.0 $ \\
$30\arcmin  $\phantom{$\backslash$ $^{\displaystyle{a_C}}$}  & $-4.5 $ & $ 0.0 $ & $ 4.6 $ & $ 4.9 $ & $ 4.9 $ & $ 4.8 $ \\
\end{tabular}

\caption{Special-relativistic angular shifts  for the optical system
described on Fig.~\ref{fig:device} rotating at $\Omega=60\ \arcsec/{\rm
s}$ and for different inclination angles $\theta = 45 \degr, 0\degr,
-45 \degr$. The mean constant shift $\delta\overline{a}_L^{\,d}$ of the
patterns due to the light propagation delays and
$\delta\overline{a}_L^{\,r}$ due to the relativistic reflection law are
given at the top of each table. Each of the tables shows the part of
the total shifts dependent on the field angles. Let us note that the
position-dependent effects in $\delta a_L^{\,d}$ and $\delta a_L^{\,r}$
have different signs and are 2-3 times larger then in the total shift
$\delta a_L=\delta a_L^{\,d}+\delta a_L^{\,r}$. On the contrary, the
effects in $\delta a_C^{\,d}$ and $\delta a_C^{\,r}$ are of the same
sign and are about 2 times less than in the sum $\delta a_C=\delta
a_C^{\,d}+\delta a_C^{\,r}$.}
\label{tab:numbers}
\end{table}

As for the one-mirror system for any value of $\theta$ the shifts due
to the light propagation delays exceed the level of 1 \muas\ and amount
to $\delta\overline{a}_L^{\,d}\sim2 \muas$. For the two-mirror system
$\delta\overline{a}_L^{\,d}$ is significantly lower than for the
one-mirror system since the effects of the motion of the primary mirror
and the motion of the focal plane largely compensate each other if just one
intermediate mirror is present.

For $\theta = 0$ the shifts due to the relativistic deflection law are
again very small as was the case for the one-mirror system. The situation with
these shifts is different for $\theta=\pm 45\degr$ where the mean shift
$\delta\overline{a}_L^{\,r}\sim0.3\,\muas$. The latter number can be
easily understood. For $\theta=\pm 45\degr$ all the light rays hit the
flat surface at an angle of about $\alpha = \pm 45\degr$ with respect
to the normal and the factor $\left|\sin\alpha\right|$ appearing in
(\ref{eq:alpha-alpha-prime}) is of the order of $1/\sqrt{2}\approx0.7$.
Each light ray of the grid hits the mirror at a slightly different
value of $\alpha$, but the main perturbation due to the relativistic
reflection law can be estimated considering the light ray going along
the optical axis. Using (\ref{eq:alpha-alpha-prime}) we obtain

\begin{eqnarray}\label{eq:delta-focal2}
\delta_2 & \simeq & 2\ \frac{v}{c}\ \frac{d_{2f}}{d_f}\,\sin^2 \theta,
\end{eqnarray}

\noindent
where $d_{2f}$ is again the distance between $\mathrm{P}_2$ and the
focal plane center as shown on Fig.~\ref{fig:device}, and $v$ is the
velocity of the point  of the mirror lying on the optical axis
($v=\Omega\,(d_{12}-r)$ for the case depicted in
Fig.~\ref{fig:device}). One can check that the mean constant shifts
$\delta\overline{a}_L^{\,r}$ as shown in Table \ref{tab:numbers} can be
recovered from (\ref{eq:delta-focal2}) almost exactly. If more flat (or
almost flat) mirrors are added, the expression can be generalized by

\begin{eqnarray}\label{eq:delta-focal}
\left| \delta_i \right| & \simeq &  \left| 2\ \frac{v_i}{c}\
\frac{d_{if}}{d_f} \ \sin \theta_i\ \sin \varphi_i\ \right|.
\end{eqnarray}

\noindent
The index $i$ is used to enumerate the surfaces along the light path, $i=1$
corresponding to the primary mirror. In our case $i=1$ is the parabolic
mirror $M_1$ and $i=2$ is the flat mirror $M_2$. The angle $\varphi_i$
is the angle between the velocity and the surface at the intersection
of the mirror $M_i$ with the optical axis applying the conventions
described on Fig.~\ref{Figure-angles}. $\theta_i$ is the angle
between the optical axis and the normal to the surface at the point of
intersection. The quantity $d_{if}$ is the distance from the center of
the focal plane $C$ to the point where the optical axis crosses the
$i$-th mirror. As defined above $d_{f}$ is the focal distance of the
optical system.

The presence of the factor $d_{if}/d_f$ in (\ref{eq:delta-focal2}) and
(\ref{eq:delta-focal}) can be explained easily: a small perturbation
$\Delta$ of the propagation direction of a light ray by a mirror
located at a distance $d_{if}$ from the focal plane causes a linear
shift on the focal plane $d_{if}\,\Delta$ which is efficiently
interpreted as an angular shift of $d_{if}/d_f\, \Delta$. In the more
general case when the intermediate reflecting surfaces are not flat,
Eq. (\ref{eq:delta-focal}) is no longer valid, but gives a reasonable
idea of the magnitude of the effect provided that all reflecting
surfaces are not too different from a flat mirror. The cumulative
effect of a series of (almost) flat mirrors will not be a direct
addition of all $\delta_i$ since the relativistic perturbation may
occur at different planes. An analytic expression in vectorial form can
be derived for the combined effect, but since the resulting formula is
rather complicated and still a rough approximation it will not be
discussed here. Eq. (\ref{eq:delta-focal}) has been checked also for
some other optical systems involving more reflecting surfaces of
different shapes, sizes and velocities. A good agreement with the
numbers from numerical ray tracing was obtained in all cases.

\section{Concluding remarks}
\label{Section:conclusion}

We have considered in detail the main relativistic effect on
the imaging by a rotating optical system which is produced by the
relativistic modification of the reflection law. We have considered two
simple optical systems containing one and two mirrors.
Although the size of the primary mirror, the focal length and the
angular velocity of rotation of both systems were defined to
agree with the corresponding parameters of Gaia, it is not clear
how big these effects will be for the real optical scheme of Gaia. We have seen
that the effects are utterly small for the one-mirror system and that
they may amount of 0.3~\muas\ for the two-mirror system. For a real
Gaia optical scheme the effect may be much larger because of the
presence of several inclined mirrors. The two examples of a rotating
optical system considered above do not allow to predict the
relativity-induced photocenter shifts for a real optical system like
Gaia. A detailed calculation of the photocenter shifts can be in
principle done using the ray tracing software developed for this
investigation.

Again the part of the effect which does not depend on the position in
the focal plane can be effectively interpreted as a constant change in
the orientation of the satellite (as discussed at the end of the
previous Section for propagation delay effects). Moreover, if a
satellite (like Gaia) has two optically different telescopes, the
difference in the main effects for these two telescopes can be
interpreted as a change in the angle between the two instruments.

In this paper we confined ourselves to ray tracing in the geometric
optics limit. A more strict way to analyze the imaging by a rotating
optical system is to apply wave optics and calculate corresponding
intensity patterns (PSF or similar characteristics). The intensity
patterns would then allow to predict the observable shifts of the
photocenters more reliably than the aberration patterns used in this
paper. Preliminary calculation with a simplified model fosters the
hope that at optical wavelengths the differences in the photocenter
shifts calculated from ray tracing and from wave optics are
negligible. However, the effects of propagation delays due to the
rotation of the telescope may play a role. This may deserve a
separate investigation.

\acknowledgement

S.K. and M.S. were partially supported by the BMWi grant 50\,QG\,0601
awarded by the Deutsche Zentrum f\"ur Luft- und Raumfahrt e.V. (DLR).


\appendix

\section{Reflection of a light ray by an arbitrarily moving mirror}
\label{Section:Reflection}

\subsection{Notation and conventions}
\label{sec:notation}

Let us first summarize the most important notation and conventions used
throughout the paper:
\begin{itemize}

\item $c$ is the velocity of light in vacuum.

\item Lowercase latin indices $a$, $i$, $j$, \dots take values $1$,
$2$, $3$ and refer to spatial components of corresponding quantities.

\item Index $0$ is used for time components.

\item Greek indices $\alpha$, $\mu$, $\nu$, \dots take values $0$, $1$,
$2$ and $3$ and refer to all space-time components of corresponding
quantities.

\item The Minkowski metric is denoted by $\eta={\rm diag}(-1,+1,+1,+1)$.

\item All latin indices are lowered and raised by means
of the unit matrix $\delta_{ij} = \delta^{ij} = {\rm diag}(1, 1, 1)$,
and therefore the disposition of such indices plays no role: $a^i =
a_i$.

\item The symbol $\varepsilon_{ijk}$ is the fully antisymmetric Levi-Civita
symbol ($\varepsilon_{123} = +1$).

\item  Repeated indices imply Einstein summation rule
irrespective of their positions (e.g., $a_i\ b_i = a_1 b_1 + a_2 b_2 +
a_3 b_3)$.

\item  The spatial components of a quantity considered as a 3-vector
are set in boldface: $\ve{a} = a^i$.

\item  The absolute value (Euclidean norm) of a 3-vector $\ve{a}$
is denoted $|\ve{a}|$ and is defined by  $|\ve{a}| = \left(a^1\,a^1 +
a^2\,a^2 + a^3\,a^3\right)^{1/2}$.

\item  The scalar product of any two 3-vectors $\ve{a}$ and $\ve{b}$ with
respect to the Euclidean metric $\delta_{ij}$
is denoted by $\ve{a} \cdot \ve{b}$
is defined by  $\ve{a} \cdot \ve{b} = \delta_{ij}a^i\,b^j = a^i\,b^i$.

\end{itemize}
Below two reference systems $(t,x^i)$ and $(T,X^a)$ will be used. To
improve readability of the formulas all quantities defined in
$x^\mu=(t,x^i)$ are denoted by small latin characters with space-time
and spatial indices taken from second parts of the Greek and Latin
alphabet, respectively ($\mu$, $\nu$, \dots, $i$, $j$, \dots). All
quantities defined in $X^\alpha=(T,X^a)$ are denoted by capital latin
characters with space-time and spatial indices taken from first parts
of the Greek and Latin alphabet, respectively ($\alpha$, $\beta$,
\dots, $a$, $b$, \dots).

\subsection{Coordinate representation of an arbitrary moving mirror}
\label{sec:mirror}

Let us consider an inertial reference system of Special Relativity
$(t,x^i)$. We define an arbitrary mirror in arbitrary motion by a
bundle of particles moving along worldlines
\begin{equation}
\label{eq:wl}
x_m^{\mu}(t; \xi,\eta) = \left(t,x_m^i(t;\xi,\eta)\right).
\end{equation}
\noindent
Here $\xi$ and $\eta$ are two parameters ``numbering'' the particles.
These parameters can be though of as some non-degenerated ``coordinate
system'' on the surface of the mirror which is described by $x_m^i(t;
\xi,\eta)$ for any fixed time $t$. On the other hand, fixing $\xi$ and
$\eta$ we fix a particle on the surface of the mirror and $x_m^i(t;
\xi,\eta)$ is the worldline of that particle in coordinates $(t,x^i)$.
Further, we assume that $x_m^i(t; \xi,\eta)$ is differentiable with
respect to all its three parameters. This means in particular that the
surface of the mirror is assumed to be smooth.

Here we do not pay attention to any physical properties of the mirror
as a ``physical body'' (elasticity, deformations, etc.). We just
consider that (\ref{eq:wl}) formally defines the position of each point
of the mirror at each moment of time. The source of information for
$x_m^{i}(t; \xi,\eta)$ for realistic mirrors and the plausibility of
these representation of an arbitrarily shaped and arbitrarily moving
mirror is discussed in Section \ref{Section:Idea} above.

Starting from (\ref{eq:wl}) it is
easy to see that for any fixed time $t$ at any fixed point of the
mirror characterized by some values of $\xi$ and $\eta$ we have two
three-dimensional vectors tangent to the surface of the mirror at the
considered point as
\begin{eqnarray}
\label{eq:l-m}
l^i&=&{\partial\over\partial\xi}\,x_m^i(t;\xi,\eta),
\\
m^i&=&{\partial\over\partial\eta}\,x_m^i(t;\xi,\eta).
\end{eqnarray}
\noindent
Then a coordinate vector normal to the surface of the mirror at that
point can be defined as
\begin{equation}
\label{n}
n^i=\varepsilon_{ijk}\,l^j\,m^k.
\end{equation}
\noindent
The order of vectors $l^i$ and $m^i$ in (\ref{n}) is arbitrary and
corresponds to a choice of the sign in the definition of $n^i$ (if
$n^i$ is a normal vector then $-n^i$ is also a normal). Not restricting
the generality we assume below that (\ref{n}) defines that $n^i$ which
is directed toward the ``working surface'' of the mirror, that is for
any incoming light ray $\sigma^i$ which hits the mirror at the
considered point one has $\ve{\sigma}\cdot\ve{n}<0$. Let us note
immediately that this normal vector $n^i$ has clearly no physical
meaning since it is defined in some arbitrary coordinate system
$(t,x^i)$. It is, however, straightforward to compute $n^i$ if
$x_m^{i}(t; \xi,\eta)$ is given. Below we show how to relate
$n^i$ to a physically meaningful normal vector at some point of the mirror
as observed by an observer instantaneously co-moving with the considered
point of the surface.

The coordinate velocity of any point of the mirror reads
\begin{equation}
\label{eq:v}
v^i_m={\partial\over\partial t}\,x_m^i(t,\xi,\eta).
\end{equation}

\subsection{Transforming the mirror surface from one
inertial reference system to another}
\label{sec:transforming}

Let us now define another reference system $(T,X^a)$ moving with
constant velocity $v^i$ with respect to $(t,x^i)$. The coordinates
$(T,X^a)$ and $(t,x^i)$ are related by a Lorentz transformation of
the form
\begin{eqnarray}
c\,t&=&\L^0_0\,c\,T+\L^0_a\,X^a ,\\
x^i&=&\L^i_0\,c\,T+\L^i_a\,X^a.
\end{eqnarray}
\noindent
The $\L$ matrix coefficients are given by
\begin{eqnarray}
\label{eq:l1}
\L^0_0 &=& \g ,
\\
\label{eq:l2}
\L^0_a &=& k^a\,\g ,
\\
\label{eq:l3}
\L^i_0 &=& k^i\,\g ,
\\
\label{eq:l4}
\L^i_a &=& \delta^{ia} + \frac{\g^2}{1+\g}\,k^i\,k^a ,
\\
\label{eq:l5}
\g &=& \left(1-\k\cdot\k \right)^{-\frac{1}{2}} ,
\\
\label{eq:l6}
\k &=& \frac{1}{c}\,\ve{v} .
\end{eqnarray}
\noindent
The inverse transformation reads
\begin{eqnarray}
c\,T&=&\Lb^0_0\,c\,t+\Lb^0_i\,x^i,
\\
X^a&=&\Lb^a_0\,c\,t+\Lb^a_i\,x^i.
\end{eqnarray}
\noindent
with
\begin{eqnarray}
\Lb^{0}_{0} &=& \g ,\label{eq:l7}\\
\Lb^{0}_{i} &=& -k^{i}\,\g ,\label{eq:l8}\\
\Lb^{a}_{0} &=& -k^{a}\,\g ,\label{eq:l9}\\
\Lb^{a}_{i} &=& \delta^{ia} + \frac{\g^{2}}{1+\g}k^{i}k^{a}.\label{eq:l10}
\end{eqnarray}

Clearly, in the reference system $(T,X^a)$ the mirror can be also represented
in the same form as in Section \ref{sec:mirror}
\begin{equation}
\label{eq:X-m}
X_m^{\alpha}(T;\xi,\eta) = \left(T,X_m^a(T;\xi,\eta)\right),
\end{equation}
\noindent
where fixed values for $\xi$ and $\eta$ should correspond to one and
the same surface particle in both coordinate systems.
The vectors tangent and normal to the surface read
\begin{eqnarray}
\label{L}
L^a&=&{\partial\over\partial\xi}\,X_m^a(T;\xi,\eta),
\\
\label{M}
M^a&=&{\partial\over\partial\eta}\,X_m^a(T;\xi,\eta),
\\
\label{N}
N^a&=&\varepsilon_{abc}\,L^b\,M^c.
\end{eqnarray}
\noindent
Here again, $N^a$ is a coordinate normal vector which has, generally speaking,
no physical meaning. The coordinate velocity of a point of the mirror is given by
\begin{equation}
\label{eq:V}
V^a_m={\partial\over\partial T}\,X_m^a(T;\xi,\eta).
\end{equation}

Let us now relate the vectors $L^a$, $M^a$, $N^a$ and $V^a_m$ to the
corresponding ones in the reference system $(t,x^i$). This this end we
consider the coordinate transformation of the events defined by
(\ref{eq:wl}) and (\ref{eq:X-m})
\begin{eqnarray}
\label{eq:t-T}
c\,T &=& \Lb^{0}_{0}\,c\,t + \Lb^{0}_{i}\,x_m^{i}(t;\xi,\eta), \\
\label{eq:x-X}
X^a_m(T;\xi,\eta) &=& \Lb^{a}_{0}\,c\,t + \Lb^{a}_{i}\,x_m^{i}(t;\xi,\eta).
\end{eqnarray}
\noindent
The function $X^a_m(T,\xi,\eta)$ is thus defined by
(\ref{eq:t-T})--(\ref{eq:x-X}) implicitly since (\ref{eq:t-T}) should
be inverted to give $t$ as function of $T$, $\xi$ and $\eta$ and that
$t$ should be substituted into (\ref{eq:x-X}) to give the explicit
dependence of $X^a_m$ on $T$, $\xi$ and $\eta$. Clearly, that inversion
cannot be done explicitly for any $x_m^{i}(t;\xi,\eta)$. However, the
partial derivatives of $X^a_m(T;\xi,\eta)$ representing $L^a$, $M^a$
and $V^a_m$  can be calculated as derivatives of an implicit function.
A straightforward algebra gives
\begin{eqnarray}
\label{eq:V-v}
V^a_m&=&c\,\frac{\Lb^a_0+\Lb^a_i\,k_m^i}{\Lb^0_0 + \Lb^0_i\,k_m^i},
\\
\label{L-l}
L^a&=&\Sb^a_i\,l^i
\\
\label{M-m}
M^a&=&\Sb^a_i\,m^i
\\
\label{Sb}
\Sb^a_i&=&\Lb^a_i-\Lb^0_i\frac{\Lb^a_0+\Lb^a_j\,k_m^j}{\Lb^0_0+\Lb^0_j\,k_m^j},
\end{eqnarray}
\noindent
or inverting
\begin{eqnarray}
\label{eq:v-V}
v^i_m&=&c\,\frac{\L^i_0+\L^i_a\,K_m^a}{\L^0_0 + \L^0_a\,K^a_m},
\\
\label{l-L}
l^i&=&S^i_a\,L^a,
\\
\label{m-M}
m^i&=&S^i_a\,M^a,
\\
\label{S}
S^i_a&=&\L^i_a-\L^0_a\frac{\L^i_0+\L^i_b\,K^b_m}{\L^0_0+\L^0_b\,K^b_m},
\end{eqnarray}
\noindent
with $k^i_m=c^{-1}\,v_m^i$ and $K^a_m=c^{-1}\,V_m^a$. Equations
(\ref{eq:V-v}) and (\ref{eq:v-V}) coincide with the law for velocities
addition in Special Relativity. One can also check by direct
calculation that $S^{i}_{a}\Sb^{a}_{j}=\delta^i_j$ and $\Sb^{a}_{i}
S^{i}_{b}=\delta^a_b$.

Using (\ref{Sb}) and (\ref{S}) one can see that
\begin{eqnarray}
\label{identity1}
\Sb^b_j\,\Sb^c_k\varepsilon_{abc}&=&{1\over \gamma\,(1-\ve{k}\cdot\ve{k}_m)}\,\,S^i_a\varepsilon_{ijk},
\\
\label{identity2}
S^j_b\,S^k_c\varepsilon_{ijk}&=&\gamma\,(1-\ve{k}\cdot\ve{k}_m)\,\Sb^a_i\varepsilon_{abc}.
\end{eqnarray}
\noindent
Now using these formulas, definitions (\ref{N}) and (\ref{n})
and relations (\ref{L-l})--(\ref{M-m}) and (\ref{l-L})-(\ref{m-M})
one can prove that $N^a$ and $n^i$ are related as
\begin{eqnarray}
\label{n-N}
N^a&=&{1\over \gamma\,(1-\ve{k}\cdot\ve{k}_m)}\, S^i_a\,n^i,
\\
\label{N-a}
n^i&=&\gamma\,(1-\ve{k}\cdot\ve{k}_m)\, \Sb^a_i\,N^a.
\end{eqnarray}
\noindent
To proof (\ref{identity1})--(\ref{identity2}) we used the identity
\begin{equation}
\label{identity3}
\varepsilon_{ajc}\,\delta^{kb}+
\varepsilon_{kac}\,\delta^{jb}+
\varepsilon_{jkc}\,\delta^{ab}=
\varepsilon_{ajk}\,\delta^{bc}.
\end{equation}

\subsection{Observable and coordinate normal vectors}
\label{Section:observable-normal}

Let us consider an infinitely small element of the mirror which is
characterized by infinitely small intervals around some fixed values of
$\xi$ and $\eta$. The velocity of the element is $v_m^i(t;\xi,\eta)$ in
the laboratory reference system $(t,x^i)$. Let us now identify the
constant velocity $v^i$ of the reference system $(T,X^a)$ relative to
$(t,x^i)$ with $v_m^i(t;\xi,\eta)$ of the considered point given by
$\xi$ and $\eta$ and at some fixed moment of time: $v^i\equiv
v_m^i(t;\xi,\eta)$. Then $(T,X^a)$ is a momentarily co-moving inertial
reference system of the considered infinitesimal element of the mirror.
The coordinates basis of $(T,X^a)$ gives an orthonormal tetrad of an
observer co-moving with the considered element of mirror.
That reference system can be used
to describe the results of instantaneous observations made by that
observer.

In particular, $N^a$ is the observable normal vector which will be used
below to formulate the reflection law for the light rays as it is seen
by the co-moving observer. From now on, $N^a$ is always used in this
sense (that is, from now on we always put $\ve{k}_m=\ve{k}$).
Normalizing the vectors one can see that the normal
unit vector $\vNun=\ve{N}/|\ve{N}|$ to the surface as seen by an
observer instantaneously co-moving with a particular point of the
mirror relates to the normal unit vector $\vnun=\ve{n}/|\ve{n}|$ seen
by an observer at rest relative to $(t,x^i)$ as
\begin{eqnarray}
\label{eq:n2N}
\vNun &=&\frac{1}{ \sqrt{1 - (\k \cdot \vnun) ^{2} } }
               \left( \vnun - (\k\cdot\vnun)\, \frac{\gamma}{1+\gamma}\,
               \k\right), \\
\label{eq:N2n}
\vnun &=&  \frac{1}{ \sqrt{1 + \g^2(\k\cdot\vNun) ^{2} } }
               \left( \vNun + (\k\cdot\vNun)\, \ggg \k\right) .
\end{eqnarray}

It is illustrative to see that this transformation of normal vectors
can be derived by the transformation rule of 4-vectors. Let us again
consider a certain surface element in its instantaneously co-moving
inertial coordinate system $(T,X^a)$. In that system we consider the
3-components of the surface normal vector $\hat\vN$ as spatial
components of the covariant 4-vector $\hat N_\alpha = (0, \hat N^a)$. A
Lorentz transformation of this 4-vector to coordinates $(t,x^i)$ leads
to result (\ref{eq:N2n}) after normalization.

\subsection{Wave vectors in the two inertial reference systems}
\label{sec:lightrays}

In order to consider the light reflection from the mirror we first need
to relate the wave vectors of the incoming and outgoing light rays in
the two considered coordinate system. In the reference system $(t,x^i)$
the incoming light ray is characterized by its null wave vector $p^\mu$
($\eta_{\mu\nu}\,p^\mu\,p^\nu=0$). The unit light ray direction
$\sigma^i$ ($\ve{\sigma}\cdot\ve{\sigma}=1$) in that reference system
is related to $p^\mu$ as $\sigma^i=p^i/p^0$. In the reference system
$(T,X^a)$ the null wave vector of the same light ray is $P^\alpha$, and
the unit light ray direction $\Sigma^a=P^a/P^0$
($\ve{\Sigma}\cdot\ve{\Sigma}=1$). The frequencies $f$ and $F$ of the
light in the corresponding reference systems are linearly proportional
to $p^0$ and $P^0$, respectively.

The wave vectors $p^\mu$ and $P^\alpha$ are related by the Lorentz
transformation
\begin{eqnarray}
\label{p-P}
P^\alpha&=&\Lb^\alpha_\mu\,p^\mu,
\\
\label{P-p}
p^\mu&=&\L^\mu_\alpha\,P^\alpha,
\end{eqnarray}
\noindent
which means that the frequencies and unit light ray directions are
related as
\begin{eqnarray}
\label{sigma-Sigma}
\Sigma^a&=&{\overline\Lambda^a_0+\overline\Lambda^a_i\,\sigma^i\over
\overline\Lambda^0_0+\overline\Lambda^0_i\,\sigma^i},
\\
\label{Sigma-sigma}
\sigma^i&=&{\Lambda^i_0+\Lambda^i_a\,\Sigma^a\over
\Lambda^0_0+\Lambda^0_a\,\Sigma^a},
\\
\label{f-F}
F&=&\left(\overline\Lambda^0_0+\overline\Lambda^0_i\,\sigma^i\right)\,f,
\\
\label{F-f}
f&=&\left(\Lambda^0_0+\Lambda^0_a\,\Sigma^a\right)\,F.
\end{eqnarray}

\subsection{Reflection as seen by an instantaneously co-moving observer}
\label{sec:reflectionLaw}

For an observer instantaneously co-moving with the element of the
mirror where the light ray is reflected the following simple reflection
law is valid (in an inertial reference system of Special Relativity for
a mirror at rest)
\begin{eqnarray}
\label{F-prime}
F^\prime&=&F,
\\
\label{Sigma-prime}
\ve{\Sigma}^\prime&=&\ve{\Sigma}-2\,(\hat{\ve{N}}\cdot\ve{\Sigma})\,\hat{\ve{N}},
\end{eqnarray}
\noindent
where $\hat{\ve{N}}$ is the observable unit normal vector to the
surface of the mirror at the point of reflection as discussed in
Section \ref{Section:observable-normal} above. The reflection law
(\ref{Sigma-prime}) means simply that the component of $\ve{\Sigma}$
perpendicular to the surface changes its sign. This is automatically
guarantees that the angle of incidence is equal to the angle of
reflection and that the incoming ray $\ve{\Sigma}$, the reflected ray
$\ve{\Sigma}^\prime$ and the normal ${\hat N}$ are coplanar. The same
equations (\ref{F-prime}) and (\ref{Sigma-prime}) are valid for,
respectively, time and space components of wave vectors before and
after reflection.

We consider this reflection law as given, but it is well known how to
derive it from Maxwell equations for electromagnetic field for a mirror
at rest \citep{Jackson:1975}. In the instantaneously co-moving
reference system $(T,X^a)$ the coordinate velocity of the reflecting
point vanishes but its acceleration may differ from zero. However, the
acceleration cannot affect the instantaneous process of reflection in
virtue of the equivalence principle as long as the conditions for
geometrical optics are satisfied (see also Section \ref{sec:gradient}
below).

\subsection{Reflection as seen by a laboratory observer}
\label{sec:final}

Now combining the reflection law in reference system $(T,X^a)$ with
the transformations discussed in Sections
\ref{sec:transforming}--\ref{sec:lightrays} one gets the reflection law
as seen in reference system $(t,x^i)$ where the mirror is arbitrarily
moving
\begin{eqnarray}
\label{f-prime}
f^\prime&=&f\,
{1+\left(\k \cdot \vnun \right)\,[\,\vnun \cdot \left(\,\k - 2\,\ve{\sigma} \right)\,]
\over 1-\left(\k \cdot \vnun \right)^2},
\\
\label{sigma-prime}
\ve{\sigma}^\prime&=&\frac{\left( 1-\left(\k \cdot \vnun \right)^2 \right) \ve{\sigma} +
              2\left( \k \cdot \vnun - \ve{\sigma} \cdot \vnun \right) \vnun}
            {1+\left( \k \cdot \vnun \right)^2 - 2(\k \cdot \vnun) \ (\ve{\sigma} \cdot \vnun) }.
\end{eqnarray}
\noindent
Here, $f^\prime$ and $\ve{\sigma}^\prime$ are the frequency and the
unit direction of the reflected light ray in the reference system
$(t,x^i)$. These expressions are valid at each point of the mirror
surface in arbitrary motion. Let us remind that $\ve{k}= \ve{v}_m/c$,
where $\ve{v}_m$ is the coordinate velocity of the reflecting point of
the mirror at the moment of reflection. Velocity $\ve{v}_m$ can be
computed from any mathematical representation of the mirror surface
(for example, from (\ref{eq:v})).

The same way can be used to derive the 4-momentum or 4-velocity of a
particle $p^{\prime \mu}$ after a completely elastic collision with a
surface of infinite mass:
\begin{eqnarray}
\label{doppler}
p^{\prime 0} &=& p^{0} - 2 \knh
              \ \left( \frac{
              \ve{p} \cdot \vnun - \knh \ p^{0}
              }{1- \left( \knh \right)^2}
              \right),
\\
\label{vector-momentum}
p^{\prime i} &=& p^i - 2 \hat{n}^i
              \ \left( \frac{
              \ve{p} \cdot \vnun - \knh \ p^{0}
              }{1- \left( \knh \right)^2}
              \right),
\end{eqnarray}
\noindent
where $p^{\mu}$ is wave vector of the particle before the collision.
Recalling the relations between wave vectors and frequencies and
directions for a photon it is easy to see that Eqs.
(\ref{doppler})--(\ref{vector-momentum}) are equivalent to
(\ref{f-prime})--(\ref{sigma-prime}).

Let us note two important properties of
(\ref{f-prime})--(\ref{sigma-prime}), also applicable to
(\ref{doppler})--(\ref{vector-momentum}):
\begin{itemize}
\item[1.]
Also in the reference system $(t,x^i)$ the reflected direction
$\ve{\sigma}^\prime$ lies in the plane defined by the incoming ray
$\ve{\sig}$ and the normal vector $\vnun$.

\item[2.]
The reflected ray is only affected by the projection of the
velocity $\ve{v}_m$ on the vector $\vnun$.
\end{itemize}
\noindent
The latter property implies that the relation between
$\ve{\sigma}^\prime$ and $\ve{\sigma}$ coincides with the usual
reflection law (\ref{Sigma-prime}) if the velocity $\ve{v}_m$ is
perpendicular to $\vnun$. This case is relevant for liquid (rotating)
mirrors and was discussed by \citet[problem 1.19]{lightman:1973},
\citet{ragazzoni:1995} and \citet{hickson:1995}. Our result (no
relativistic effects on reflection law in that case) coincides with
that of \citet{lightman:1973} and \citet{hickson:1995}.

Multiplying both sides of (\ref{sigma-prime}) by $\vnun$ and use the
following definitions for the angles between vectors
\begin{eqnarray}
-{\ve{\sig}}\cdot\vnun &=&\cos\alpha, \label{eq:ang1}\\
\ve{\sigma}^\prime \cdot \vnun &=& \cos \alpha^\prime, \label{eq:ang2} \\
\k \cdot \vnun &=& k \cos\left(\varphi - \frac{\pi}{2}\right) = k\,\sin \varphi,  \label{eq:ang3}\
\end{eqnarray}
\noindent
($k=|\ve{k}|=|\ve{v}_m|/c$) one obtains a relation between the angle of
incidence $\alpha$ and angle of reflection $\alpha^\prime$
\begin{eqnarray}
\label{f-prime-a}
f^\prime&=&f\,{1+2\,k\,\sin\varphi\,\cos\alpha+k^2\,\sin^2\varphi
\over 1-k^2\,\sin^2\varphi},
\\
\label{eq:aleksandar}
\cos \alpha^\prime &=& \frac{ 2\, k\, \sin \varphi +
                      \left(1 + k^2\,\sin^2 \varphi\right)\, \cos \a
                  }
                  {1 + k^2 \sin^2 \varphi
                     + 2\,k\,\sin \varphi \cos \a
                  }\ .
\end{eqnarray}
\noindent
The latter equation can be also re-written into an equation relating $\sin\alpha$
and $\sin\alpha^\prime$:
\begin{eqnarray}
\label{eq:aleksandar:sin}
\sin \alpha^\prime &=& \sin\alpha\,
{1-k^2\,\sin^2\varphi \over 1+2\,k\,\sin\varphi\,\cos\alpha+k^2\,\sin^2\varphi}.
\end{eqnarray}
\noindent
Comparing (\ref{f-prime-a}) and (\ref{eq:aleksandar:sin}) one can see that
$f\sin\alpha=f^\prime\,\sin \alpha^\prime$.

\begin{figure}[htb]
\centering
\includegraphics[width=8.0cm]{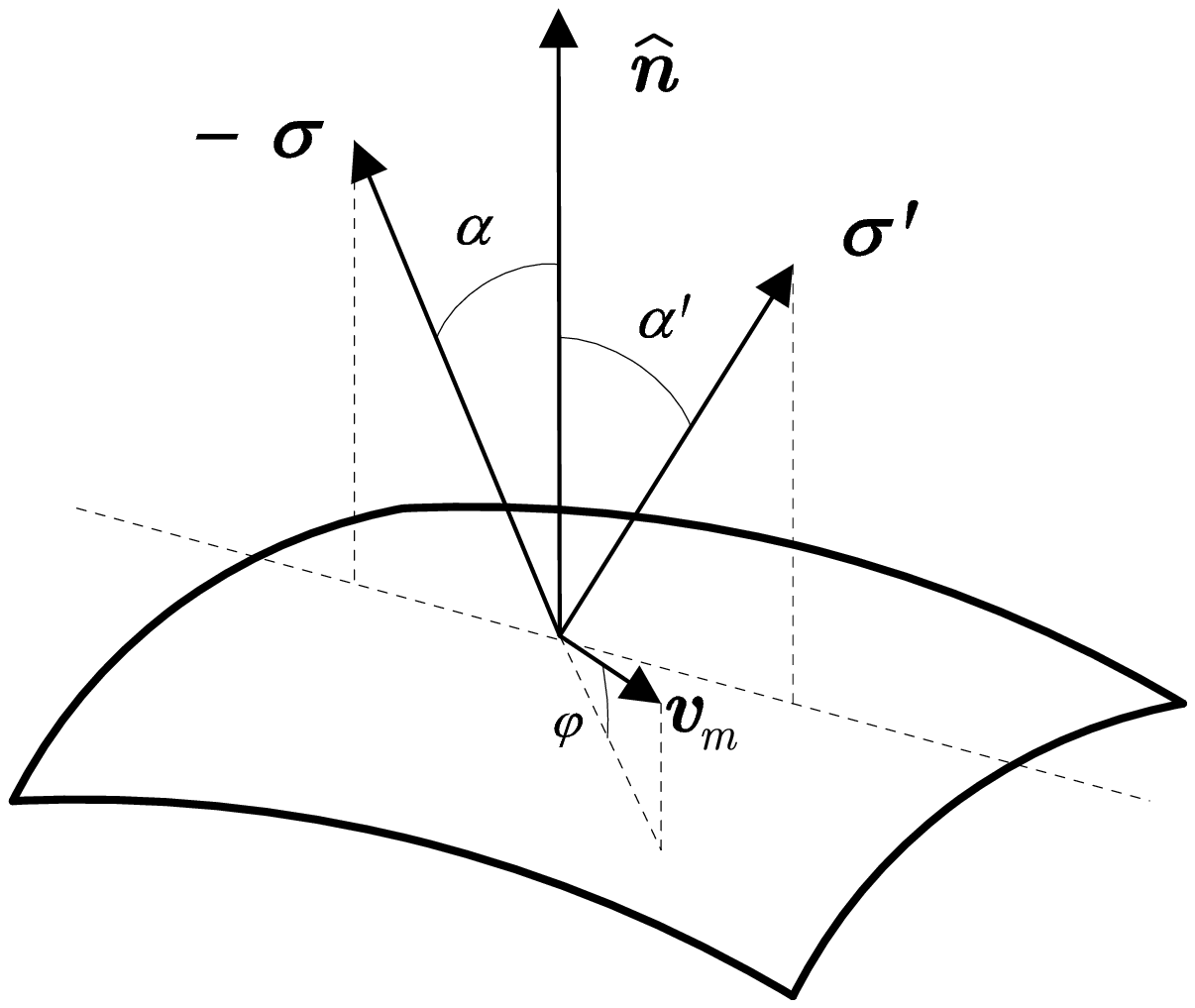}
\caption{Vectors and angles at the point of reflection. Vector $\vnun$
is a coordinate unit vector perpendicular in the Euclidean sense to the
surface of the mirror at the reflection point. Vectors $\ve{\sigma}$
and $\ve{\sigma}^\prime$ are unit directions of propagation of the
incoming and reflected light rays, respectively. Vector $\ve{v}_m$ is
the velocity of the point of the mirror at which the reflection occurs.
Angle $\alpha$ between the direction $-\ve{\sigma}$ toward the source
of the incoming light ray and vector $\vnun$ is $0\le\alpha\le\pi/2$.
Angle $\alpha^\prime$ between the propagation direction
$\ve{\sigma}^\prime$ of the reflected light ray and vector $\vnun$ is
again $0\le\alpha^\prime\le\pi/2$. Finally $\varphi$
($-\pi/2\le\varphi\le\pi/2$) is the angle between the velocity vector
$\ve{v}_m$ and the plane tangential to the mirror at the point of
reflection. The latter angle is negative if the angle between
$\ve{v}_m$ and $\vnun$ is greater than $\pi/2$ and positive otherwise.
Because of the special-relativistic effects angle $\alpha$ and
$\alpha^\prime$ are in general different.
}
\label{Figure-angles}
\end{figure}

Angles $\alpha$, $\alpha^\prime$ and $\varphi$ are illustrated on
Fig.~\ref{Figure-angles}. The angle $\alpha$ lies between 0 and $\pi/2$
(since we always consider that the incoming light ray comes to the
mirror from one particular side of the tangent plane to the mirror's
surface at the point of reflection). For the same reason we have
$0\le\alpha^\prime\le\pi/2$. Angle $\varphi$ lies between $-\pi/2$ and
$\pi/2$. It is negative if the angle between $\k$ and $\vnun$ is
greater than $\pi/2$ and positive otherwise.

\subsection{Particular case of a flat mirror moving with a constant velocity}

As a particular example let us apply the developed scheme to a flat
mirror moving at constant velocity in reference frame $(t,x^i)$. The
mathematical expression for that is a worldline equation
(\ref{eq:wl}) in the form
\begin{equation}
\label{flat-mirror-x}
\ve{x}_m(t,\xi,\eta)=\ve{x}_{m0}+\ve{l}\,\xi+\ve{m}\,\eta+\ve{v}_m\,t,
\end{equation}
\noindent
where $\ve{l}$, $\ve{m}$, $\ve{v}_m$ and $\ve{x}_{m0}$ are constant vectors
defining position, velocity and orientation of the mirror. It is easy to see that
in coordinates $(T,X^a)$ one gets
\begin{equation}
\label{flat-mirror-X}
\ve{X}_m(T,\xi,\eta)=\ve{X}_{m0}+\ve{L}\,\xi+\ve{M}\,\eta+\ve{V}_m\,T,
\end{equation}
\noindent
where vectors $\ve{V}_m$, $\ve{L}$ and $\ve{M}$ are related to
$\ve{v}_m$, $\ve{l}$ and $\ve{m}$ by (\ref{eq:V-v}), (\ref{L-l}) and
(\ref{M-m}), respectively, and $X^a_{m0}=\Sb^a_i\,x^i_{m0}$. Eq.
(\ref{flat-mirror-X}) implies that a flat surface remains flat in any
inertial reference system.

Since for a flat mirror $\ve{l}$ and $\ve{m}$ are constants, the unit
normal vector $\hat{\ve{n}}$ is also a constant. Since $\ve{v}_m$ is
also time-independent the same reflection law described by
(\ref{sigma-prime}) or (\ref{eq:aleksandar}) is valid for any point of
the mirror and at any moment of time. One can check that
(\ref{eq:aleksandar}) coincides with the results of
\citet{aleksandar:2004}. We believe, however, that our framework is
more general than that of \citet{aleksandar:2004} since we do not {\it
assume} the vectors $\ve{\sigma}$, $\vnun$, $\k$ and
$\ve{\sigma}^\prime$ to be coplanar, and our derivation is valid for an
arbitrary mirror in arbitrary motion.

Note that the central result of \citet{aleksandar:2004} coincides
with the formula derived by \cite{Einstein:1905} in the particular
case of a flat mirror moving with constant velocity directed
perpendicular to the surface when $\sin\varphi=1$ (see also
\citet[problem 1.18]{lightman:1973}). \citet{bolotovskii:1989} have
derived the same relation as \cite{Einstein:1905} by solving Maxwell
field equations directly in the coordinates where the mirror is
moving.

\subsection{Low velocity limit}\label{sec:low}

It is useful to derive the first-order expansion of
(\ref{f-prime})--(\ref{eq:aleksandar}) in powers of $v_m/c$ since in
practice the velocity of the mirror will be small compared to the light
velocity. One gets
\begin{eqnarray}
\label{f-prime-low}
f^\prime&=&f\,\left(1-2\,(\ve{\sigma}\,\hat{\ve{n}})\,(\k\cdot\hat{\ve{n}})+\OO2\right),
\\
\label{sigma-prime-low}
\ve{\sigma}^\prime&=&\ve{\sigma}-2\,(\ve{\sigma} \cdot \vnun)\,\ve{\nun}
\nonumber \\
&&
+2\,(\k\cdot\vnun)\,\left[\left(1-2\,(\ve{\sigma}\cdot\vnun)^2\right)\,\vnun+(\ve{\sigma}\cdot\vnun)\,\ve{\sigma}\right]
\nonumber \\
&&
+\OO2,
\end{eqnarray}
\noindent
or
\begin{eqnarray}
\label{f-prime-low-angles}
f^\prime&=&f\,\left(1+2\,k\,\sin\varphi\,\cos\alpha+\OO2\right),
\\
\label{sigma-prime-low-angles}
\cos\alpha^\prime&=&\cos\alpha+2\,k\,\sin\varphi\,\sin^2\alpha+\OO2,
\\
\label{sigma-prime-low-angles-sin}
\sin\alpha^\prime&=&\sin\alpha-k\,\sin\varphi\,\sin2\alpha+\OO2.
\end{eqnarray}
\noindent
The first two terms in the right-hand side of (\ref{sigma-prime-low})
represent just the usual reflection law and the rest contains the
largest relativistic effects. Eq. (\ref{sigma-prime-low-angles}) shows
that
\begin{equation}\label{eq:alpha-alpha-prime}
\alpha^\prime-\alpha=-2\,k\,\sin\varphi\,\sin\alpha+\OO2.
\end{equation}
\noindent
This expression can be used to estimate the difference
$\alpha^\prime-\alpha$ for many realistic situations.

\subsection{Derivation of results by means of Maxwell's equations}
\label{sec:gradient}

It is illustrative to see how the results
(\ref{f-prime-a})--(\ref{eq:aleksandar}) can be derived directly
from Maxwell's equations. It is well known that the usual reflection
law can be obtained from Maxwell's theory by a principle of phase
matching: the phase of the incoming wave $\Phi$ should agree with
the phase of the outgoing wave $\Phi^\prime$ on the mirror
surface $m$ (e.g., \citet[Section 7.3]{Jackson:1975}):
\begin{equation}
\label{phase-matching-eq}
\left.\Phi \right|_m = \left.\Phi^\prime \right|_m \, .
\end{equation}
\noindent
The central results (\ref{f-prime-a})--(\ref{eq:aleksandar}) can simply
be derived from the principle of phase matching in case of a flat
mirror moving with constant velocity $\ve{v}_m$ with respect to
inertial coordinates $x^\mu$ where the observer is at rest. The mirror
$x^i_m$ is given in this case by (\ref{flat-mirror-x}).
The constant (time- and position-independent) unit normal vector is
again denoted as $\vnun$.
Maxwell's equations in inertial coordinates lead to the the usual wave equation
of the form
\begin{equation}
- \left( {1 \over c^2}{\partial^2 \over \partial
t^2} + \Delta \right) \Psi = 0
\end{equation}
\noindent
that is solved, e.g., by a monochromatic plane wave of the form
\begin{equation}
\Psi = a\,\exp\left({i\,p_\mu\,x^\mu}\right) \equiv a\,\exp\left(i\,\Phi\right)
\end{equation}
\noindent
with a wave vector $p^\mu = (p^0, \ve{p})$ satisfying the usual
null condition
\begin{equation}\label{null-cond}
-p^0\,p^0 + \ve{p} \cdot \ve{p} = 0 \, .
\end{equation}

The principle of phase matching (\ref{phase-matching-eq})
then determines both the law of reflection and the Doppler shifts of
``photon'' frequencies. Let us decompose the wave vector $\ve{p}$ into
a tangential and a normal part with respect to the surface normal:
\begin{eqnarray}
\ve{p} &=& \ve{p}_T + p_n\, \vnun,
\\
\ve{p}_T &=&\vnun\times(\ve{p}\times\vnun),
\\
p_n &=&\ve{p}\cdot\vnun.
\end{eqnarray}
\noindent
Then phase matching on the mirror surface leads to
\begin{eqnarray}
\ve{p}_T&=& \ve{p}^\prime_T \\
p^0 - {1\over c}\,\ve{v}_m\cdot\vnun\ p_n &=& p^{\prime0} - {1\over c}\,\ve{v}_m\cdot\vnun\ p_n^\prime \, .
\end{eqnarray}
\noindent
or using the null condition the two matching equations for frequencies
$f$ and $f^\prime$ and direction angles $\alpha$ and $\alpha^\prime$
(see Fig.~\ref{Figure-angles}) take the form
\begin{eqnarray}
f\,\sin \alpha &=&
f^\prime\,\sin \alpha^\prime \\
f\,\left(1 + {1\over c}\,\ve{v}_m\cdot\vnun\,\cos \alpha \right) &=&
f^\prime\,\left(1 - {1\over c}\,\ve{v}_m\cdot\vnun\,\cos \alpha^\prime \right)\, .
\end{eqnarray}
\noindent
Straightforward algebra then leads to the results (\ref{f-prime-a}) and
(\ref{eq:aleksandar}) above. Note, that this phase-matching argument
works in a simple way for plane mirrors and plane waves that
mathematically are infinitely extended both in space and time. Such a
treatment, however, is meaningful for any mirror as long as the
conditions for geometrical optics are satisfied, i.e., as long as
amplitude, polarization and wave vector do not change significantly
over a  distance determined by the wavelength. This implies that the
acceleration $a_m$ of the mirror should satisfy a constrain of the form
$a_m\ll c^2/\lambda$, where $\lambda$ is the wavelength of the
radiation.

\end{document}